\documentclass[a4paper,11pt]{article}

\usepackage{hyperref}
\usepackage[not1,notextcomp,lcgreekalpha]{stix}
\usepackage{authblk}
\usepackage{url}
\usepackage{textcomp}
\usepackage{enumerate}
\usepackage{xcolor}
\usepackage{booktabs}
\usepackage{array}

\usepackage{geometry}
\providecommand{\keywords}[1]{\textbf{\textit{Keywords: }} #1}

\usepackage{amsmath}
\usepackage{cases}
\usepackage[pagewise]{lineno}
\usepackage{graphicx}
\usepackage{caption}
\usepackage[labelfont={bf,footnotesize},textfont=footnotesize]{subcaption}
\usepackage{float}
\usepackage{esint}
\usepackage{mathtools}
\usepackage{listings}
\newcommand{\class}[1]{\textbf{\itshape #1}}
\newcommand{\method}[1]{\texttt{#1}}

\newcommand{\dee}{\mathrm{d}}
\newcommand{\trp}{\mathrm{T}}
\newcommand{\broomstyx}{BROOMStyx}

\DeclareMathOperator{\trace}{\mathrm{tr}}

\title{\bfseries A new object-oriented framework for solving multiphysics problems via combination of different numerical methods}
\author[1,2]{Juan Michael Sargado\thanks{jusa@norceresearch.no}}

\affil[1]{\small Department of Mathematics, University of Bergen, All\'egaten 41, 5007 Bergen, Norway}
\affil[2]{\small NORCE Norwegian Research Centre AS, Nyg{\aa}rdsgaten 112, 5008 Bergen, Norway}
\date{}

\begin{document}
\maketitle
\begin{abstract}
	\noindent Many interesting phenomena are characterized by the complex interaction of different physical processes, each often best modeled numerically via a specific approach. In this paper, we present the design and implementation of an object-oriented framework for performing multiphysics simulations that allows for the monolithic coupling of different numerical schemes. In contrast, most of the currently available simulation tools are tailored towards a specific numerical model, so that one must resort to coupling different codes externally based on operator splitting. The current framework has been developed following the C++11 standard, and its main aim is to provide an environment that affords enough flexibility for developers to implement complex models while at the same time giving end users a maximum amount of control over finer details of the simulation without having to write additional code. The main challenges towards realizing these objectives are discussed in the paper, together with the manner in which they are addressed. Along with core objects representing the framework skeleton, we present the various polymorphic classes that may be utilized by developers to implement new formulations, material models and solution algorithms. The code architecture is designed to allow achievement of the aforementioned functionalities with a minimum level of inheritance in order to improve the learning curve for programmers who are not acquainted with the software. Key capabilities of the framework are demonstrated via the solution of numerical examples dealing on composite torsion, Biot poroelasticity (featuring a combined finite element-finite volume formulation), and brittle crack propagation using a phase-field approach. \\[10pt]
\end{abstract}

\section{Introduction}
Computer simulations involving multiple interacting physical processes are becoming more and more common, in large part due to the availability of machines with fast processors having multiple cores and high memory capabilities. While the most demanding simulations still need to be run on high-performance computing clusters, nowadays a majority of simulations related to both research and industry are in fact carried out on much more modest systems such as desktop and even laptops computers. Likewise, the field of computational science has seen rapid expansion. In 1928 when Courant, Friedrich and Lewy first described in the now famous CFL stability condition, finite differences had yet to find wide application for solving partial differential equations. In contrast, there is now a plethora of available techniques dealing on numerical solution of PDEs. For instance, finite elements have become the method of choice for solid mechanics applications, while in computational fluid dynamics finite difference and finite volume methods find popular usage, and to some extent also boundary elements. More recently, a substantial amount of research has gone towards the development of particle and meshfree schemes, and the invention of new approaches such as peridynamics, isogeometric analysis and virtual elements.

The main ingredient in the performance of scientific computing is of course, software. Initial impetus for the finite element method originated in the aerospace industry with Boeing, and likewise advancement of FE software was driven by industrial needs beginning with the work of Wilson at Aerojet and the subsequent development of the NASTRAN\textsuperscript{\textregistered} code in the late 1960s.\cite{Clough2004,Wilson1993} The commercial FE codes Abaqus\textsuperscript{\textregistered} and ANSYS\textsuperscript{\textregistered} had their initial releases in the 1970s and remain among the most popular simulation tools in industry. On the other hand, the present landscape of computational science is markedly different. For instance it can be argued that there is now a clearer divide between academia and industry, with most of the programming work related to implementation of new approaches and models being done by academic researchers utilizing interpreted languages such as MATLAB and Python. The popularity of these platforms stems from the fact that they allow for rapid implementation, prototyping and visualization of results as well easier debugging due to access to intermediate states of variables during execution time. On the other hand, codes used to generate results in publications are often hand-tailored to the specific problems being solved and are impossible to apply without substantial modification to other cases. The unfortunate result is that a lot of different numerical methods are accompanied by implementations that are not robust enough for general testing, which in turn hinders their investigation and acceptance by the community at large. Additionally, execution speed becomes a critical factor for simulations dealing with large problem sizes, in which case it is advisable to make use of optimized software written in a compiled language such as C/C++ or Fortran.

In recent years, the trend has been towards open-source software packaged as libraries in order to provide the greatest amount of flexibility to researchers. One such project that has found quite a bit of success in the community is deal.II\cite{Bangerth2007}, which is written in C++ and designed for performing numerical simulations based on adaptive quadrilateral/hexahedral meshes with automatic handling of hanging nodes. Another is FEniCS\cite{Logg2012}, a software platform written in C++/Python for implementing finite element solutions to general partial differential equations that gives emphasis to the proper setup of function spaces and variational forms. On the other hand, the DUNE project\cite{Bastian2003} was initially designed as a collection of modules providing modern interfaces to various legacy codes and on which libraries implementing numerical methods can be built, for instance FE and discontinuous Galerkin\cite{Dedner2010}. In general, a developer implementing some particular physical model makes calls to said libraries via a main file. The latter is usually written in the same language as used by the library, or in some cases via a special interpreted language such as UFL\cite{Alnaes2014}. Furthermore the standard procedure is for problems involving the same physical model but different geometries and boundary conditions to be handled by separate main files. In effect, developers (e.g.\ researchers) who perform the coding and compilation for their respective models are also the intended end users of the resulting executable binaries.

An alternative approach is to write code that is implemented as components within existing software. This is known as the framework approach and is the option offered by most commercial simulation software (e.g. Abaqus and ANSYS) along with some open-source codes such as OOFEM \cite{Patzak2001}, OpenFOAM \cite{Weller1998} and FEAP \cite{Taylor2014}. An important characteristic of the framework approach is \emph{inversion of control}: program flow is defined and controlled by the existing framework, with developers writing code that is designed to be called by the framework itself at specific instances during program execution. An immediate consequence of the framework approach is the clear delineation between the role of coder and end user. In particular, an implicit goal is for end users to be able to solve different problems (i.e. in terms of geometry and boundary/initial conditions) without having to recompile the software. While a framework's more rigid structure allows for less flexibility compared to what can be achieved through a library approach, it nevertheless accomplishes two things: first, it forces a developer to consider an existing flow of control when conceptualizing and implementing a particular model and discourages the use of procedures that are too complicated to implement efficiently. At the least, it provides a common ground on which to judge the performance penalty incurred by such algorithms in relation to other methods. Secondly, developers are forced to follow the general procedure built into the framework with regards to the interaction between software and end user (for instance concerning specification of model parameters), which makes the deployment and testing of new models and methods straightforward for those already familiar with the style of input utilized by the software. 

At present, most available simulation codes (both proprietary and open-source) are designed to accommodate a specific method such as FEM, or are otherwise tailored towards certain applications. Creating a general software package that can straightforwardly implement different numerical schemes is challenging, as these methods generally utilize varying formulations of the governing equations, and likewise may require different ways of imposing boundary conditions depending on the approximation properties of the assumed function space.

In this paper, we present the design of {\broomstyx}, a new open-source software framework that seeks to address the challenge of attaining seamless coupling between fundamentally dissimilar formulations coming from different contributors. In particular, a main goal of the software to allow for monolithic solution strategies for combined formulations, or in the case of operator splitting schemes, to avoid the writing of intermediate output files that is necessary when coupling together different codes. The remainder of this paper is structured as follows: Section \ref{sec:designConsiderations} explains the motivation behind development of the framework as well as some of its key features. Section \ref{sec:frameworkArchitecture} deals with the code architecture and aspects arising from its object-oriented design; different groups of classes that make up the framework are identified, along with the role of each class type and how it interacts with other components of the software. Section \ref{sec:linearAlgebra} expounds on the implementation of custom container classes for real-valued vector and matrices as well as operator overloading in a way that allows for user-friendly syntax in the coding of vector/matrix operations without sacrificing efficiency of computation. Meanwhile, aspects related to shared memory parallelization are briefly discussed in Section \ref{sec:parallelization}. To show that the current framework is applicable to a wide class of problems, we solve several numerical examples dealing on various topics in the mechanics of solids and porous media, and which have been specifically chosen to demonstrate novel features of the software as pointed out in previous sections. These can be found in Section \ref{sec:examples}, and include a discussion of the relevant theory for each problem and important implementation details in addition to the actual numerical results. Finally, concluding remarks are given in Section \ref{sec:conclusion}.

\section{Design considerations} \label{sec:designConsiderations}
The name {\broomstyx} is short for ``Broad Range, Object-Oriented Multiphysics Software'', and as stated previously the project aims to provide a sufficiently flexible and extensible general framework for developers to implement new models and combine different formulations while at the same time giving end users a powerful tool for performing sophisticated numerical simulations. In this section, we discuss the main points that guided development of the {\broomstyx} code, and some notable design features that distinguish it from currently available software.

\subsection{Equal focus on developers and end users}
Apart from core programmers of the software who are responsible for maintenance of the general code structure and the incorporation of fundamental modifications/additions to existing functionality, the persons interacting with numerical simulation software can be seen as belonging to at least one of two groups. In the first group are contributing developers, i.e. people involved with model creation and the development of solution algorithms who see the code as a platform upon which to build and test new models and methods. The other group consists of end users, essentially non-programmers who are interested in using the existing functionality in the software to analyze problems that occur in real life. Equal focus on both groups means that the overall usefulness of the project depends on the ease with which one can
\begin{enumerate}[a)]
	\item implement new formulations, material models and solution algorithms without having to achieve mastery of the entire software code, and
	\item make use of such models in real-life applications (e.g.\ with complex geometries, multiple components and large numbers of unknowns) without having to write additional lines of code.
\end{enumerate}
In {\broomstyx}, this is addressed first of all through a framework-type design that gives rise to a fixed workflow for end users with regard to the setup of input (both involving problem specifics and the domain discretization) and also the handling of simulation results for subsequent output writing and visualization. In particular we do not attempt to re-implement preliminary mesh construction or visualization of results. Instead it is preferred to implement functionality that converts data stored using the internal structure of the code to something that usable by third party software such as Gmsh\cite{Geuzaine2009} and Paraview\cite{Ayachit2015}. Likewise, the object-oriented paradigm enables setting of a definite structure with regard to publicly accessible methods in base classes so that subsequent contributions implementing new models and methods in the form of derived classes are guaranteed to be compatible with other components of the framework.
 
Most currently available modeling software favor a target audience that consists either of researchers who are more concerned with basic method development or very specialized scientific modeling, or of industry practitioners who do not necessarily have a very deep knowledge of theory but nonetheless should be able to perform analysis of real world problems without having to get into the finer details of the formulation beyond providing the problem geometry, material properties and boundary/initial conditions. Good software design that bridges the gap between these two very different groups can potentially have tremendous impact as it gives practitioners access to the latest models and techniques. Furthermore it provides a means for these new developments to be studied rigorously in terms of their applicability to real world scenarios. Needless to say, the open-source nature of such software is vital. While some proprietary codes possess the flexibility of accommodating custom implementations (and indeed software such as COMSOL\textsuperscript{\textregistered} are built for exactly this purpose), oftentimes it is not justifiable from a cost perspective to purchase new software simply to test some novel formulation or algorithm that is yet to be proven suitable for widespread usage.

\subsection{Generality of equations and formulations}
One of the core objectives of {\broomstyx} is to allow native coupling of different numerical methods/formulations within a single framework. Consequently, the code structure is not specialized for any one specific numerical method. Nevertheless a universal aspect shared by these methods is the transformation of some given set of partial differential equations (PDEs) into a system of algebraic equations involving unknown degrees of freedom (DOFs). Let $\mathbf{u}$ be a vector of scalar field variables and their spatial derivative components up to some order, i.e.
\begin{linenomath}
\begin{equation}
	\mathbf{u} = \left\{ u_1, \; u_2, \ldots, \; \partial_i u_1, \; \partial_i u_2, \ldots, \partial_i \partial_j u_1, \; \partial_i \partial_j u_2, \ldots \right\}
\end{equation}
\end{linenomath}
in which $\partial_k$ signifies a derivative along the $k$-th dimension. Using this representation, a system of PDEs can be written in the general form
\begin{linenomath}
\begin{equation}
	\frac{\partial^2 }{\partial t^2} \mathbf{A} \left( \mathbf{u}, \mathbf{x}, t \right) + \frac{\partial}{\partial t} \mathbf{B} \left( \mathbf{u}, \mathbf{x}, t \right) + \mathbf{C} \left( \mathbf{u}, \mathbf{x}, t \right) = \mathbf{D} \left( \mathbf{x}, t \right).
	\label{eq:genEq}
\end{equation}
\end{linenomath}
The code operates by evaluating the global discrete approximations for $\mathbf{A}$, $\mathbf{B}$, $\mathbf{C}$ and $\mathbf{D}$ along with suitable approximations for temporal derivatives, and enforcing the equality of left and right hand sides over some given time interval. When dealing with solution algorithms that require assembly of a global coefficient matrix from local contributions, no assumptions are made regarding the structure of local matrices. In saddle point problems for instance where local matrices have the form
\begin{linenomath}
\begin{equation}
	\mathbf{K} = \left[ \begin{array}{cc} \mathbf{K_{11}} & \mathbf{K_{12}} \\[4pt] \mathbf{K_{21}} & \mathbf{0} \end{array} \right],
\end{equation}
\end{linenomath}
it should be possible exclude partitions consisting of zeros when constructing the sparsity profile of the global matrix and during subsequent assembly in order to avoid unnecessarily increasing the global matrix density.

Furthermore, {\broomstyx} aims to place no restriction on the kinds of equations that can be modeled, other than that numerical quantities be real-valued. In particular, the code should be able to handle models that make use of an arbitrary number and combination of different DOF types without having to alter the core structure of the software. This creates a challenge with regard to the assignment of DOFs, since it is not possible to anticipate beforehand all the different DOF types that a particular model might require. The solution adopted in {\broomstyx} is to split the task of managing DOF types between developers and end users. That is, specification of the required number of DOF types is provided by the model implementation, while the actual assignment is done by the end user at runtime, who specifies both the number of DOF types to be initialized per geometric entity (i.e., node, cell or facet) and how these are to be utilized by the model, as demonstrated in Section \ref{sec:examples}.

\subsection{Model coupling across domains}
\begin{figure}
	\centering
	\begin{subfigure}[b]{0.3\textwidth}
		\centering
		\includegraphics[width=0.9\textwidth]{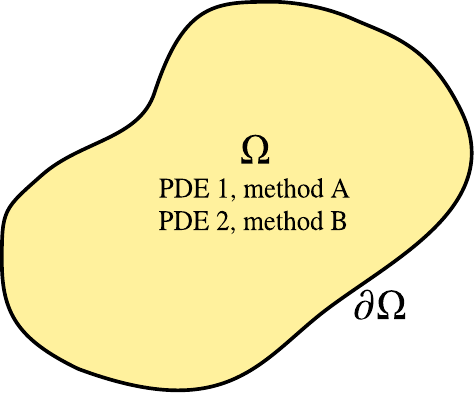}
		\caption{}
		\label{fig:domainConfig1}
	\end{subfigure}
	\begin{subfigure}[b]{0.33\textwidth}
		\centering
		\includegraphics[width=0.9\textwidth]{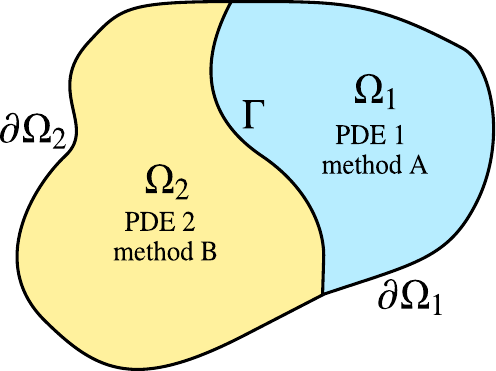}
		\caption{}
		\label{fig:domainConfig2}
	\end{subfigure}
	\hfill
	\begin{subfigure}[b]{0.33\textwidth}
		\centering
		\includegraphics[width=\textwidth]{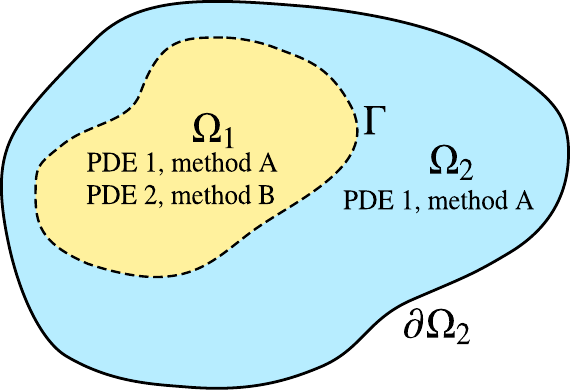}
		\caption{}
		\label{fig:domainConfig3}
	\end{subfigure}
	\caption{Different domain configurations for a multiphysics problem consisting of two PDEs.}
	\label{fig:domainConfigs}
\end{figure}

To illustrate some of the complexities that must be addressed when designing user-friendly and efficient multiphysics code, consider the case of a problem involving two coupled PDEs. There are several possible domain configurations that might arise in connection with such a problem, some of which are shown in \ref{fig:domainConfigs}. For illustrative purposes, we further assume that the aforementioned equations are approximated using different numerical methods.
Figure \ref{fig:domainConfig1} represents the basic scenario most often encountered in multiphysics modeling where component PDEs share the same domain $\Omega$ and boundary $\partial\Omega$. On the other hand, Figure \ref{fig:domainConfig2} illustrates a case where the two equations live on non-overlapping domains that share a common interface, $\Gamma$. A classic example of such a scenario occurs in fluid-structure interaction problems, where $\Gamma$ acts as a boundary for both $\Omega_1$ and $\Omega_2$. An additional rule must be specified relating the boundary conditions of the different PDEs, and depending on modeling choices this can result in either one-way or two-way coupling of the equations. On the other hand Figure \ref{fig:domainConfig3} can be understood as an extension of Figure \ref{fig:domainConfig1} in that one of the PDEs is defined over $\Omega_1 \cup \Omega_2$ and the other over $\Omega_1$. Consequently, the latter must have its boundary conditions defined on $\Gamma$. In contrast, this interface is non-existent with respect to the other equation whose boundary conditions are defined on $\partial\Omega_2$. In order for the software to be effective, the end user must have the possibility to set up these different configurations solely via the input file.

\subsection{Information access via manager classes}
In traditional object-oriented programming, a member object generally does not have access to the methods of its parent or other objects higher up in the compositional hierarchy. Rigid adherence to this principle means that all information needed by a class implementing some particular numerical model must be passed as arguments when calling a particular class method, which is disadvantageous for two main reasons. First of all, such an approach makes method calls extremely long. The same phenomenon is obtained when a user-implemented function is not given access to other methods/functions within the framework. This is evident for instance in the definition of user-programmed element (UEL) subroutines in Abaqus, shown below\cite{AbaqusManual}:
\begin{lstlisting}[language=fortran,literate={'}{'}1]
      SUBROUTINE UEL(RHS,AMATRX,SVARS,ENERGY,NDOFEL,NRHS,NSVARS,
     1 PROPS,NPROPS,COORDS,MCRD,NNODE,U,DU,V,A,JTYPE,TIME,DTIME,
     2 KSTEP,KINC,JELEM,PARAMS,NDLOAD,JDLTYP,ADLMAG,PREDEF,NPREDF,
     3 LFLAGS,MLVARX,DDLMAG,MDLOAD,PNEWDT,JPROPS,NJPROP,PERIOD)
C
      INCLUDE 'ABA_PARAM.INC'
C
      DIMENSION RHS(MLVARX,*),AMATRX(NDOFEL,NDOFEL),PROPS(*),
     1 SVARS(*),ENERGY(8),COORDS(MCRD,NNODE),U(NDOFEL),
     2 DU(MLVARX,*),V(NDOFEL),A(NDOFEL),TIME(2),PARAMS(*),
     3 JDLTYP(MDLOAD,*),ADLMAG(MDLOAD,*),DDLMAG(MDLOAD,*),
     4 PREDEF(2,NPREDF,NNODE),LFLAGS(*),JPROPS(*)
     ...
     ...
      RETURN
      END
\end{lstlisting}
More importantly, the amount of information accessible to model/algorithm developers becomes hard-coded in the declaration of base classes and cannot be subsequently modified without altering each and every derived class already coded. To circumvent this problem, {\broomstyx} makes use of manager classes, whose public methods are accessible to all objects. While this may seem to break encapsulation, it endows model-implementing classes with almost unlimited flexibility, and at the same time allows them to access only the information that is actually required by the model at a given instant. In this sense, the software can be viewed as a library/framework hybrid that aims to retain the benefits of both approaches.
\section{Framework architecture} \label{sec:frameworkArchitecture}
\begin{figure}
	\includegraphics[width=\textwidth]{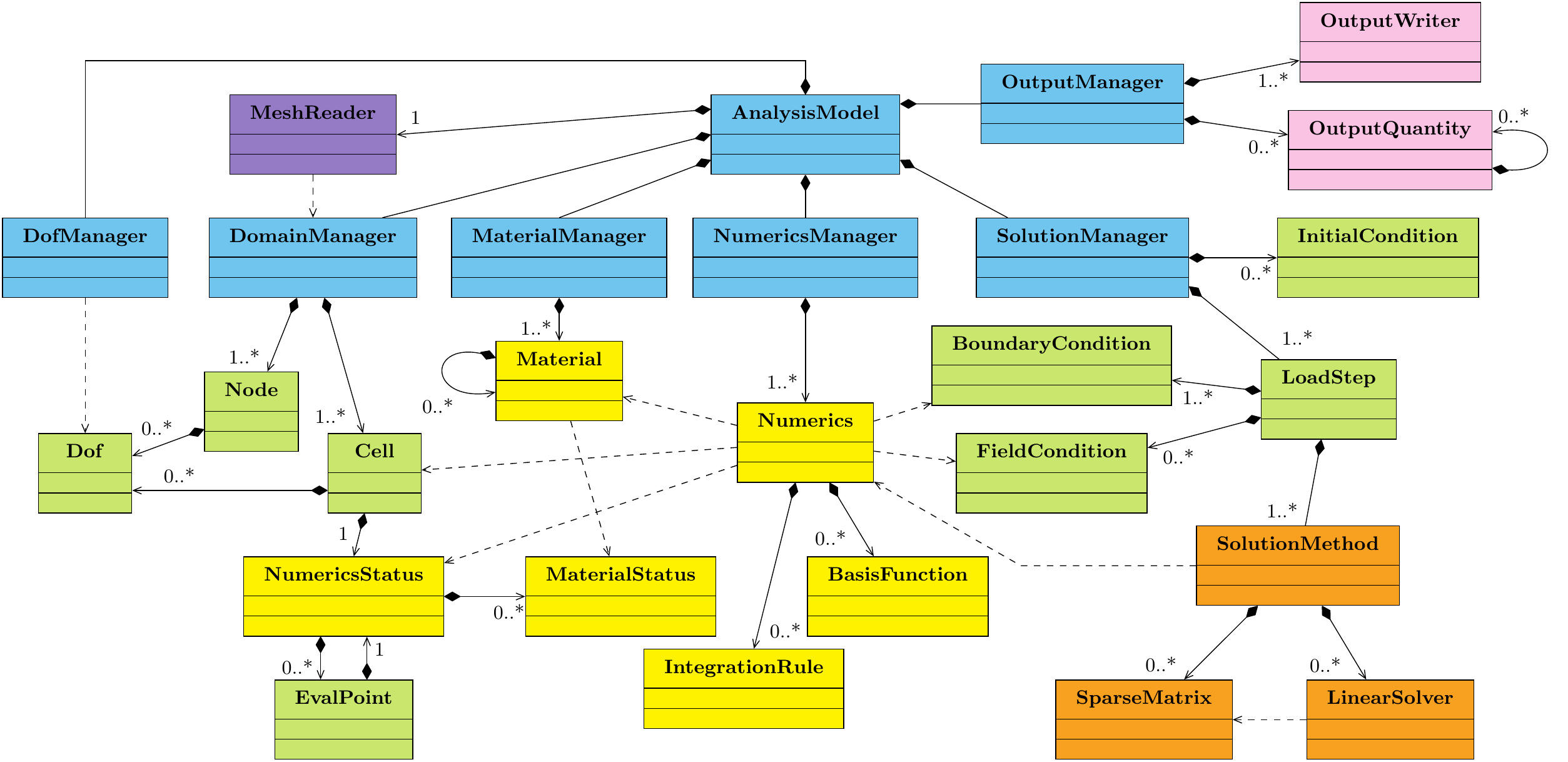}
	\caption{Class diagram showing the general structure of the BROOMStyx code. 
	}
	\label{fig:frameworkArchitecture}
\end{figure}

The general structure of the {\broomstyx} code is comprised of different class groups as shown in Figure \ref{fig:frameworkArchitecture}.
The first group consists of non-polymorphic classes within the \texttt{broomstyx} namespace whose modification/maintenance is left to the core developers of the framework. These are referred to as core classes, and may further be divided into two subgroups depending on the number of actual objects that can be instantiated from them at runtime. The rest are polymorphic classes which deal with different aspects of model and algorithm development, the reading of different mesh formats and the writing of output.

\subsection{Single-instance core classes}
Single-instance core classes consist of \class{AnalysisModel}, \class{ObjectFactory}, and the various manager classes which are colored blue in Figure \ref{fig:frameworkArchitecture}.  \class{AnalysisModel} provides two public methods. The first is \method{initializeYourself()} which instantiates the manager classes, reads the input file and creates all the necessary objects for the simulation through calls to the various methods provided by said manager classes. The second is \method{solveYourself()} which carries out the actual simulation. To ensure that only a single instance is generated for the entire duration of the program, \class{AnalysisModel} is implemented as a Meyers singleton pattern. In addition, the various manager classes have private constructors and destructors that are accessible to \class{AnalysisModel} by virtue of being declared a friend class. Finally, \class{AnalysisModel} is responsible for instantiating the proper \class{MeshReader} object to read the mesh file supplied by the user.

Among the manager classes, \class{DomainManager} is responsible for the management of objects corresponding to the nodes and cells of the mesh, setting up information regarding connectivity, and providing methods for accessing the degrees of freedom associated with each geometric entity. On the other hand, the creation and deletion of the actual degrees of freedom are handled by the \class{DofManager} class, which also provides methods to set subsystem, group and equation numbers at each DOF, and to update and finalize values for the unknown variables of the simulation. In addition, \class{DofManager} provides a method for reading multi-freedom constraints from the input file which can then be used to set up constraints involving master/slave DOFs. 

The \class{MaterialManager} class is responsible for reading the number and types of materials that are declared in the input file, and for instantiating and initializing the corresponding \class{Material} objects. In general, destruction of managed objects is performed when the destructor for their corresponding manager object is invoked; this is true for the \class{MaterialManager} class as well as the subsequent manager classes that will be discussed. Furthermore in the input file each material must be assigned a unique integer label, and the manager class provides a method to access instantiated materials based on their assigned labels. Class \class{NumericsManager} is similar to \class{MaterialManager} in its function, i.e. it reads the different numerics types specified in the input file together with their assigned integer label, and instantiates the corresponding \class{Numerics} objects. Likewise, it provides a method for accessing the instantiated numerics based on their assigned labels.

The main role of the \class{SolutionManager} class is to read the number of load steps from the input file and instantiate the corresponding \class{LoadStep} objects. It also reads all specified initial conditions from the same file, and imposes them at the beginning of the solution process. In addition, it is responsible for instantiating and managing \class{UserFunction} objects that implement user-programmed functions. Finally, the \class{OutputManager} class is in charge of managing all things related to simulation output. It instantiates a particular \class{OutputWriter} object according to the type specified by the user in the input file. It also reads the different output quantities that need to be calculated and creates the appropriate \class{OutputQuantity} objects. The manager class provides methods for initializing these objects as well as for calculating and writing output at the end of each converged time step. 

Finally, {\broomstyx} contains an \class{ObjectFactory} class, which as its name implies is in charge of the actual creation of objects based on derived classes. To accomplish this, the framework provides the macro
\begin{verbatim}
    registerBroomstyxObject(<baseClass>, <derivedClass>)
\end{verbatim}
within the \texttt{broomstyx} namespace, which must be invoked in the source code of each derived class. Like \class{AnalysisModel}, the \class{ObjectFactory} class is implemented as a Meyers singleton and provides two sets of methods. The first set involves registration of derived classes that is accessed indirectly via the aforementioned macro, while the second set creates objects of the different derived class types. The various manager classes instantiate derived objects via a call to \class{ObjectFactory} which then returns a pointer to the instantiated object, at which point responsibility for the said object is transferred to the calling manager class. At program startup, a map is automatically created within the \class{ObjectFactory} instance whose key entries are the names of all available derived classes, which are in turn paired with pointers to the corresponding class constructors.

As mentioned in Section \ref{sec:designConsiderations}, public methods of the manager classes can be accessed anywhere within the \texttt{broomstyx} namespace and are initiated via a call to the \class{AnalysisModel} object which has global scope. For instance, one can retrieve the nodes associated with some particular cell object via the following call:
\begin{lstlisting}
std::vector<Node*> cellNodes = analysisModel().domainManager().giveNodesOf(targetCell);
\end{lstlisting} 

\subsection{Multiple-instance core classes}
The subgroup of multiple-instance core classes consist of class types pertaining to geometrical entities and also concepts related to the solution process. Class \class{Dof} is an abstraction for a single degree of freedom, and it stores data pertaining to which DOF group, solution stage and subsystem it belongs, its equation number, current and converged values of its associated unknown and the corresponding residual value. It also keeps track of its status, i.e., whether it is constrained (and if so, the value of the constraint) and in the case that it is of slave type, the pointer to its associated master DOF. The class itself does not provide any methods for accessing its members, which are all declared private. Instead their retrieval and update is done through \class{DofManager} that is declared a friend class of \class{Dof}. Meanwhile, the \class{Node} class is an abstraction for a geometric entity, specifically a point in 3-dimensional space. It stores the point's coordinates, a unique ID number, a vector of pointers to \class{Dof} objects that live on the node, and a vector of real numbers whose length is set by the end user at runtime and which serves to store values that will subsequently be written into output files. Furthermore it contains two sets of pointers to \class{Cell} objects that are attached to the node, one for domain cells and the other for boundary cells.

The \class{Cell} class on the other hand has a more nuanced function. Rather than simply being an abstraction for a particular geometric entity, its primary purpose is to act as an anchor to which \class{Numerics} objects can attach their associated data and perform the necessary calculations related to the governing equations. Every \class{Cell} object contains a vector of real numbers, the length of which is set at runtime in order to store values needed for writing output similar to the case of the \class{Node} class. Each \class{Cell} object also stores its own type (e.g., 3-node triangle, 10-node tetrahedron, general $n$-point polygon) and dimension, its ID number, physical entity, and a vector of pointers to \class{Dof} objects associated with the cell. In addition it stores data related to connectivity, such as its neighbors and faces. Every \class{Cell} instance also contains a vector of pointers to \class{Node} objects, whose exact relation to the cell may vary depending on the particular formulation being used. For instance within the context of a standard finite difference scheme each cell may be associated to some FD stencil so that the cell nodes are meant to coincide with the stencil points, whereas in a finite element scheme the \class{Cell} object can be seen as representing a single element, and the nodes as entities to which shape functions are associated in order to approximate a piecewise solution. \class{Cell} objects are also used to represent geometrical entities such as surfaces on which boundary and interface conditions are applied. Lastly, when cell faces are required to be constructed by a particular formulation, these are also represented as \class{Cell} objects. For this reason the \class{DomainManager} class distinguishes between three types of cells: a) domain cells that are associated with some given \class{Numerics} instance\footnote{This group may also include lower-dimensional interface cells, for example when the latter are associated with a \class{Numerics} object implementing some inter-domain or multiphysics coupling.}, b) boundary cells that are associated some particular boundary condition, and c) face cells.

The \class{EvalPoint} class is an abstraction for a single evaluation point. It stores the coordinates of its location in addition to its assigned weight (in the case of Gauss integration), and contains a pointer to a \class{NumericsStatus} object that in turn stores data such as history variables required by \class{Numerics} and \class{Material} objects for performing their calculations. As shown in Figure \ref{fig:frameworkArchitecture}, \class{EvalPoint} objects typically occur as components of some \class{NumericsStatus} instance. Nonetheless their existence is not mandatory; cells which have only one evaluation point for example can have the necessary numerics and material variables stored directly as members of the immediate \class{NumericsStatus} object connected to the cell instance, thus avoiding compositions of the type \texttt{Cell->NumericsStatus->EvalPoint->NumericsStatus}.

The \class{BoundaryCondition} class is used to store information regarding a single boundary condition, such as the physical entity label corresponding to the boundary on which the condition is to be applied, the particular BC type, the target \class{Dof} type and the implementing numerics. This last piece of information is necessary since a \class{BoundaryCondition} object is not in charge of actually imposing its given BC. Rather, said object calculates the value (possibly time-dependent) of the relevant quantity at the required location on the boundary and leaves the imposition of this value to the implementing \class{Numerics} object.\footnote{The main reason for adopting such a strategy is that numerical methods generally differ in the way they impose boundary conditions. For instance some schemes such as the Element-Free Galerkin method\cite{Belytschko1994} make use of shape functions that do not possess Kronecker-delta properties, hence Dirichlet conditions must be imposed differently compared to finite element schemes even if both are based on a variational formulation of the governing equations. Passing on the actual implementation of boundary conditions and the like to the specific numerics allows us to avoid having to overload \class{BoundaryCondition} for each new \class{Numerics} derived class.} The \class{FieldCondition} and \class{InitialCondition} classes are have similar function --- the former is used to describe domain-wide quantities, for example specific source terms and possibly also parameters of governing equations that are made to vary in space and time, whereas the latter returns the value of a specific initial condition at some given location within the domain. As with \class{BoundaryCondition}, the actual implementations of these field and initial conditions are left to the respective associated \class{Numerics} objects.

Class \class{LoadStep} is an abstraction for a solution step taken over a range of time, which in turn can be further divided into substeps. Essentially the workhorse class of the entire code, \class{LoadStep} objects are responsible for marching the solution forward in time according parameters specified by the end user. Each \class{LoadStep} object reads its required data from the input file; these consist of the starting time and ending time of the solution step, the initial time increment for each substep, the maximum allowed number of substeps, all the boundary and field conditions to be imposed for the current solution step, the particular type of solution method to be used, and all special pre-/post-processing procedures available in numerics implementations that are required to be performed according to end user specifications. The \class{LoadStep} class provides the method \method{solveYourself()}, which carries out the sequence procedures listed in Table \ref{tab:loadStepSolnProcedure}.
\begin{table}
	\caption{General solution procedure executed on invocation of \texttt{LoadStep->solveYourself()}.}
	\label{tab:loadStepSolnProcedure}
	\centering
	\begin{tabular}{lp{0.7\textwidth}}
		\toprule
		1. & Initialize solvers to be used by solution methods. \\
		2. & Carry out pre-processing routines. \\
		3. & Find constrained and active degrees of freedom. \\
		4. & Determine target time for current substep. If this exceeds the specified end time, shorten it so that the target time exactly coincides with the end time. \\
		5. & For each solution stage, have the corresponding \class{SolutionMethod} object calculate the solution given the specified boundary and field conditions. \\
		6. & If solution in step 5 converged, finalize the relevant data. Otherwise, implement contingency procedures (e.g., change the time-increment and go back to step 4, or alternatively terminate the simulation due to unconverged results.) \\
		7. & Update current time and possibly also the time increment (adaptive time stepping). If end time has been reached, then proceed to step 8. Otherwise, go back to step 4. \\
		8. & Carry out post-processing routines. \\ 
		\bottomrule
	\end{tabular}
\end{table}
Note that steps 2 and 8, pre- and post-processing do not refer to meshing/visualization procedures as commonly understood in FEM terminology, but instead denote access points built into the framework which can be exploited by model developers to perform calculations that are not meant for execution during assembly and solution of system equations, for instance the initialization of some history variables based on the latest converged solution.

\subsection{Polymorphic classes}
The {\broomstyx} framework contains several polymorphic classes that can be used to extend functionality of the software. As illustrated in Figure \ref{fig:frameworkArchitecture}, these fall into four groups. The first group is composed of the \class{MeshReader} class and deals with the interpretation of mesh files generated from third party software and the creation of the corresponding entities in the internal memory of the software. Meanwhile, the second group is concerned with model creation, and consists of the \class{Numerics}, \class{Material}, \class{ShapeFunction} and \class{IntegrationRule} classes. The third group allows for the implementation of different solution schemes and back-end solvers, and includes the \class{SolutionMethod}, \class{LinearSolver} and \class{SparseMatrix} classes. The last group allows the production of different output formats and the calculation of specialized quantities through the classes \class{OutputWriter} and \class{OutputQuantity}.

\subsubsection{MeshReader}
The \class{MeshReader} class provides a way for the {\broomstyx} framework to understand output files from external CAD and mesh generation tools which contain the discretization (i.e. nodes and elements) to be used in the numerical simulation. At present, the framework allows for the instantiation of one \class{MeshReader}-derived object that reads a single mesh file containing all the necessary components of the discretization, whose name is specified in the input file. At the same time, said \class{MeshReader} object interfaces with the \class{DomainManager} object to create the resulting geometric entities in memory.

\subsubsection{Numerics}
The \class{Numerics} class is an abstraction for a PDE (or system of PDEs) together with its numerical discretization, and is the fundamental building block around which the rest of the code was designed. It  contains methods for calculating local contributions corresponding to different components of the governing equations as described in Section \ref{sec:designConsiderations}. Its derived classes are responsible for declaring the required geometric cell type to be used with said class and the number of DOFs that will be utilized at each node, cell or face. The base class declaration is given in Listing \ref{lst:numericsListing}, where for brevity only the virtual methods are listed.
\begin{lstlisting}[float,caption={Partial declaration of the \class{Numerics} base class, showing the methods that may be overloaded in derived classes.},label={lst:numericsListing}]
class Numerics
{
	friend class NumericsManager;	
public:
	Numerics();
	virtual ~Numerics();
	...
	virtual void initializeMaterialsAt( Cell* targetCell );
	virtual bool performAdditionalConvergenceCheckAt( Cell* targetCell, int stage );
	virtual void performPreIterationOperationsAt( int stage, int iterNum );
	virtual void printPostIterationMessage( int stage );
	virtual void readAdditionalDataFrom( FILE* fp );
	virtual void removeConstraintsOn( Cell* targetCell );
	
	virtual void finalizeDataAt( Cell* targetCell ) = 0;
	virtual void deleteNumericsAt( Cell* targetCell ) = 0;
	virtual void initializeNumericsAt( Cell* targetCell ) = 0;

	virtual std::tuple< std::vector<Dof*>, std::vector<Dof*>, RealVector >
		giveStaticCoefficientMatrixAt( Cell* targetCell, int stage, int subsys, const TimeData& time );
	virtual std::tuple< std::vector<Dof*>, RealVector >
		giveStaticLeftHandSideAt( Cell* targetCell, int stage, int subsys, const TimeData& time );
	virtual std::tuple< std::vector<Dof*>, std::vector<Dof*>, RealVector >
		giveTransientCoefficientMatrixAt( Cell* targetCell, int stage, int subsys, const TimeData& time );
	virtual std::tuple< std::vector<Dof*>, RealVector >
		giveTransientLeftHandSideAt( Cell* targetCell, int stage, int subsys, const TimeData& time,
		    ValueType valType );
	virtual void imposeConstraintAt( Cell* targetCell, int stage, const BoundaryCondition& bndCond,
		    const TimeData& time );
	
	// Error-generating virtual functions (must be implemented in derived class when called)
	virtual double giveCellFieldValueAt( Cell* targetCell, int fieldNum );
	virtual RealVector giveCellNodeFieldValuesAt( Cell* targetCell, int fieldNum );
	virtual std::vector<RealVector> giveEvaluationPointsFor( Cell* targetCell );
	virtual std::tuple< RealVector,RealVector > giveFieldOutputAt( Cell* targetCell, const std::string& fieldTag );
	virtual RealVector giveNumericsParameter( const std::string& paramTag );
	virtual std::tuple< std::vector<Dof*>, RealVector >
		giveStaticRightHandSideAt( Cell* targetCell, int stage, int subsys, const BoundaryCondition& bndCond,
		    const TimeData& time );
	virtual std::tuple< std::vector<Dof*>, RealVector >
		giveStaticRightHandSideAt( Cell* targetCell, int stage, int subsys, const FieldCondition& fldCond,
		    const TimeData& time );
		
	virtual void imposeInitialConditionAt( Cell* targetCell, const InitialCondition& initCond );	
	virtual void performPostprocessingAt( Cell* targetCell, std::string tag );
	virtual void performPrefinalizationCalculationsAt( Cell* targetCell );
	virtual void performPreprocessingAt( Cell* targetCell, std::string tag );
	virtual void setDofStagesAt( Cell* targetCell );
	...
};
\end{lstlisting}
Each instantiated \class{Numerics} object reads its required parameters from the input file, and manages its own data at each computational cell. Because different physical processes vary in the amount and types of data that they require to be stored, an auxiliary class \class{NumericsStatus} is provided whose role is to be a base container that can be specialized to suit the various \class{Numerics}-derived classes. Each computational cell includes a pointer to one \class{NumericsStatus} object in its attributes, which is instantiated by the \class{Numerics} object assigned to the cell when data storage beyond standard DOF solutions is required. As can be seen from Listing  \ref{lst:numericsListing}, the class contains methods for calculating output quantities, and for performing pre-/postprocessing of data as well as carrying out inter-iteration operations. This is to allow a maximum amount of flexibility to developers, since in BROOMStyx flow of control does not rest with any instantiated \class{Numerics} object but rather with the framework itself that in turn passes it to the specific solution method chosen by the end user at runtime. During assembly of the discrete equations, the latter cycles over all cells in the computational domain and interfaces directly with the \class{Numerics} object assigned to each cell, meaning that entities such as integration points are hidden from solution algorithms. This is a deliberate design choice, since the notion of integration points does not always make sense, for example when dealing with finite differences or control volume schemes. Hence for numerical methods based on weak formulations, the \class{Numerics} object itself is in charge of internally performing the numerical integration when calculating local matrices and vectors.

The convention adopted in BROOMStyx for stating method declarations is to clearly separate input arguments from output quantities, except when a particular variable acts as both input and output. Multiple return values are passed as a \emph{tuple}, for instance in methods that return local matrix and vector contributions. As can be observed in Listing \ref{lst:numericsListing}, matrix output is returned using something akin to coordinate (COO) format, i.e.\ as a collection of three vectors, the first two containing pointers to \class{Dof} objects which store their equation numbers (corresponding respectively to row and column indices in the global coefficient matrix) while the last vector contains the actual entries of the matrix. A key advantage of such an approach is that no assumption is made on the shape of such matrix, but rather that the method only returns local components that are actually nonzero. This is important in terms of efficiency, since the sparsity profiles for global coefficient matrices are determined by an initial assembly process that disregards local matrix values.

The following two details are worth mentioning with regard to interface design:
\begin{enumerate}[a)]
	\item The virtual methods defined in the \class{Numerics} base class for calculating matrix and vector quantities corresponding to different components of governing equations are not \emph{pure} virtual methods but rather are implemented in the base class as empty functions. For example, transient ($\partial / \partial t$) terms may be calculated with the following two methods:
	\begin{lstlisting}
virtual std::tuple< std::vector<Dof*>, std::vector<Dof*>, RealVector > 
	giveTransientCoefficientMatrixAt(Cell* targetCell, int stage, int subsys, const TimeData& time);

virtual std::tuple< std::vector<Dof*>, RealVector >
	giveTransientLeftHandSideAt(Cell* targetCell, int stage, int subsys, const TimeData& time, ValueType valType);
	\end{lstlisting}
	The existence of an empty implementation in the base class means that developers only need to override methods that are actually relevant to the problem at hand. This reduces the  amount of boilerplate code the must be written and also ensures that static equations are correctly solved when using solution methods that account for transient terms. Such functionality is necessary due to the fact that one can have a fully coupled system that features both static and time-varying PDEs.
	\item The return values for the above methods are in the form of \emph{tuples}, which must be constructed utilizing both \texttt{std::make\_tuple(...)} and \texttt{std::move} in order to avoid unwanted copying, e.g.
	\begin{lstlisting}
std::vector<Dof*> rowDof, colDof;
RealVector coefVal;
...
return std::make_tuple(std::move(rowDof), std::move(colDof), std::move(coefVal));
	\end{lstlisting}
\end{enumerate}

\subsubsection{Material}
Class \class{Material} represents the base class for constitutive models, whose role is to evaluate quantities that determine the coefficients/parameters of the governing equations. It provides methods to calculate scalar potentials, constitutive forces and tangent moduli. The class interface design is shown in Listing \ref{lst:materialListing}, and is based on the concept of a scalar potential function and its derivatives. That is, we assume that the material model can be described by some scalar potential
\begin{linenomath}
\begin{equation}
	f = f \left( \mathbf{\epsilon}, \mathbf{\alpha} \right) \text{ (material potential)}
	\label{eq:materialPotential}
\end{equation}
\end{linenomath}
whose value can be computed give the current constitutive state $\mathbf{\epsilon}$ and set of history variables $\mathbf{\alpha}$.
\begin{lstlisting}[float,caption={Declarations for \class{MaterialStatus} and \class{Material} classes.},
label={lst:materialListing}]
class Material
{
public:
	Material();
	virtual ~Material();
	
	virtual MaterialStatus* createMaterialStatus();
	virtual void destroy( MaterialStatus*& matStatus );
	virtual void initialize();
	virtual void readParamatersFrom( FILE* fp );
	virtual void updateStatusFrom( const RealVector& conState, MaterialStatus* matStatus );
	virtual void updateStatusFrom( const RealVector& conState, MaterialStatus* matStatus,
	    const std::string& label );

	// Error-generating virtual methods (must be implemented in derived class when called)
	virtual double givePotentialFrom( const RealVector& conState, const MaterialStatus* matStatus );
	virtual RealVector giveForceFrom( const RealVector& conState, const MaterialStatus* matStatus );
	virtual RealVector giveForceFrom( const RealVector& conState, const MaterialStatus* matStatus,
	    const std::string& label );
	virtual RealMatrix giveModulusFrom( const RealVector& conState, const MaterialStatus* matStatus );
	virtual RealMatrix giveModulusFrom( const RealVector& conState, const MaterialStatus* matStatus,
	    const std::string& label );
	virtual double giveMaterialVariable( const std::string& label, const MaterialStatus* matStatus );
	virtual double giveParameter( const std::string& label );

protected:
	std::string _name;
	void error_unimplemented( const std::string& method );
};
\end{lstlisting}
Likewise, it is assumed that its gradients with respect to the constitutive state variables may also be obtained as
\begin{linenomath}
\begin{align}
	\mathbf{\sigma} &= \frac{\partial}{\partial\mathbf{\epsilon}} f \left( \mathbf{\epsilon}, \mathbf{\alpha} \right) \text{ (material force)} \label{eq:materialForce} \\
	\mathbb{D} &= \frac{\partial}{\partial\mathbf{\epsilon}} \mathbf{\sigma} \left( \mathbf{\epsilon}, \mathbf{\alpha} \right) \text{ (material modulus)} \label{eq:materialModulus}
\end{align}
\end{linenomath}
Subsequently, lines 16--22 in Listing \ref{lst:materialListing} are abstractions for the calculations needed to obtain the expressions given above, and are declared \texttt{virtual} as they are to be overridden in derived class definitions. In general however, material models do not necessarily have the well defined structure implied by \eqref{eq:materialPotential}--\eqref{eq:materialModulus}, and the framework does not require that all the aforementioned methods be defined. In order to reduce boilerplate code in derived classes, these methods are predefined in the base class implementation with the default behavior of throwing an exception when called. Such feature is intended to prevent simulations from accidentally using incorrect values for material quantities due to an unintentional failure in the part of a developer to implement the proper overriding method.\footnote{If the returned quantity is a vector or matrix that is immediately used in a following calculation, then such an oversight might result in a dimension mismatch between operands. However it can also be that the nature of subsequent operations is such that no further error occurs, and the program continues execution leading to incorrect results.} An obvious limitation of the class design is that only one type of vector and matrix quantity may be returned by a particular \class{Material} instance. The reason for this is to keep interfaces simple; for more complex models one can extend functionality of the derived class by having its instance contain other \class{Material} objects in its attributes.

\subsubsection{SolutionMethod}
The \class{SolutionMethod} class is an abstraction for the solution scheme used to solve the discretized governing equations. Different algorithms can be implemented through derived classes, for instance linear static or transient schemes, Newton-Raphson, Runge-Kutta methods and so on. The base class declaration is given in Listing \ref{lst:solnMethodListing}, and includes a method for imposing constraints in order to find and mark constrained DOFs which are then excluded from the unknowns to be solved in subsequent steps.
\begin{lstlisting}[float,caption={Definition of the \class{SolutionMethod} class.},
	label=lst:solnMethodListing]
class SolutionMethod
{
public:
	SolutionMethod();
	virtual ~SolutionMethod();

	void getCurrentLoadStep();
	void imposeConstraintsAt( int stage, const std::vector<BoundaryCondition>& bndCond, const TimeData& time );
	
	virtual int  computeSolutionFor( int stage
	                               , const std::vector<BoundaryCondition>& bndCond
	                               , const std::vector<FieldCondition>& fldCond
	                               , const TimeData& time ) = 0;
	
	virtual void formSparsityProfileForStage( int stage ) = 0;
	virtual void initializeSolvers() = 0;
	virtual void readDataFromFile( FILE* fp ) = 0;

protected:
	LoadStep* _loadStep;
	std::string _name;
	
	bool checkConvergenceOfNumericsAt( int stage );
};
\end{lstlisting}
The main method of the class is \method{computeSolutionFor(...)}, which essentially performs the global equation assembly and subsequent solution via calls to the specified linear solvers in the input file. In order to perform said assembly, the sparsity profile of global systems must first be determined. This is implemented in method \method{formSparsityProfileForStage(int stage)}, which also assigns equation numbers to all unknowns. A notable feature of the {\broomstyx} framework is that it allows for multi-\emph{stage} in combination with multi-\emph{subsystem} setups. The former is essentially an implementation of a hard-coded sequential solve, in which all quantities belonging to a previous stage are updated and finalized before proceeding to the next solution stage. On the other hand multiple subsystems simply implies a partitioning of the global system of equations associated with a given stage. For example in some solution algorithms, the global system may be partitioned into several subsystems that are solved in sequence within an iterative scheme. In this case the method makes use of the DOF group assignment to determine which degrees of freedom belong to their respective subsystems. It then assigns equation numbers for the unknowns in each subsystem and determines the sparsity profiles of the associated global matrices. This must be implemented separately for each new class deriving directly from \class{SolutionMethod}, as some algorithms may need to include additional unknowns such as Lagrange multipliers, for instance in the case of arc-length methods\cite{Riks1979}. All derived classes must read their required data from the input file. This typically includes the linear solvers to be used, convergence criteria for each of the active DOF groups in the case of nonlinear solution schemes, and data related to special procedures such as line search options when these are implemented in the derived class.

\subsubsection{LinearSolver and SparseMatrix}
{\broomstyx} relies on external software packages for the solution of linear systems. Class \class{LinearSolver} is a base class for wrappers to different solver libraries in order to provide a common interface that can be utilized by other classes within {\broomstyx}. The base class declaration is given in Listing \ref{lst:LinearSolverListing}, and contains a varied number of methods that are relevant for either direct or iterative solvers.
\begin{lstlisting}[float,caption={Definitions for the \class{SparseMatrix} and \class{LinearSolver} classes.},label=lst:LinearSolverListing]
class SparseMatrix
{
public:
	SparseMatrix();
	virtual ~SparseMatrix();
	
	std::tuple< int,int > giveMatrixDimensions();
	int     giveNumberOfNonzeros();
	bool    isSymmetric();
	void    setSymmetryTo( bool true_or_false );
	
	virtual void addToComponent( int rowNum, int colNum, double val ) = 0;
	virtual void atomicAddToComponent( int rowNum, int colNum, double val ) = 0;
	virtual void finalizeProfile() = 0;
	virtual std::tuple< int*,int* > giveProfileArrays() = 0;
	virtual double* giveValArray() = 0;
	virtual void insertNonzeroComponentAt( int rowIdx, int colIdx ) = 0;
	virtual void initializeProfile( int dim1, int dim2 ) = 0;
	virtual void initializeValues() = 0;
	virtual RealVector lumpRows() = 0;
	virtual void printTo( FILE* fp, int n ) = 0;
	virtual RealVector times( const RealVector& x ) = 0;

protected:
	int _dim1;
	int _dim2;
	int _nnz;
	bool _symFlag;
};

class LinearSolver
{
public:
	LinearSolver();
	virtual ~LinearSolver();
	
	virtual void allocateInternalMemoryFor( SparseMatrix* coefMat );
	virtual RealVector backSubstitute( SparseMatrix* coefMat, RealVector& rhs );
	virtual void factorize( SparseMatrix* coefMat );
	virtual bool giveSymmetryOption();
	virtual void initialize();
	virtual void clearInternalMemory();
	virtual void setInitialGuessTo( RealVector& initGuess );
	virtual bool takesInitialGuess();
	
	virtual std::string giveRequiredMatrixFormat() = 0;
	virtual void        readDataFrom( FILE* fp ) = 0;
	virtual RealVector  solve( SparseMatrix* coefMat, RealVector& rhs ) = 0;
};
\end{lstlisting}
Derived classes must be able to specify the format of global coefficient matrices to be used during assembly of equations, read all data required by the back-end solver such as tolerances, and also perform the solution of a given linear system by internally making the necessary calls to the actual solver. At present, {\broomstyx} can interface with different versions of the direct solver PARDISO\cite{Verbosio2017,IntelMKL} as well as iterative solvers from the ViennaCL library\cite{Rupp2016} compiled to run on graphics processors via CUDA. Meanwhile, the \class{SparseMatrix} class is used to accommodate different matrix formats that may be utilized by the various linear solver packages, such as compressed sparse row (CSR) and coordinate (COO) formats. In order to allocate memory for a given spare matrix, its sparsity profile must first be determined. The class method \method{initializeProfile(int m, int n)} is used to specify the matrix dimensions, and also resets its nonzero components to the empty set. On the other hand, the method \method{insertNonzeroComponentAt(int i, int j)} declares the $\left( i,j \right)$-th component of the matrix to be nonzero. The sparsity profile of the global matrix can then be constructed by constructing the local contributions from each cell and then calling the aforementioned method for each nonzero matrix component in the local contribution, with its corresponding global indices as arguments. Finally, \method{finalizeProfile()} constructs the actual arrays for storing location indices and values according to the specified matrix format.

\subsubsection{OutputWriter and OutputQuantity}
The primary function of the \class{OutputWriter} class is to create output files containing user-chosen simulation results at certain points in time as specified in the input file. These normally consist of the values for primary unknowns at nodes or elements, as well as physical variables such as stresses, strains and so on. In general, \class{OutputWriter} objects only have access to values specified in connection with the \texttt{NodalField} and \texttt{CellFieldOutput} keywords in the input file. Derived classes are in charge of writing said output files in a format that can be read by third party visualization packages. The specific details on what to include in these files is read by the \class{OutputWriter} instance from the input file, while the frequency of creating output files relative to the start and end times of the simulation is read by \class{LoadStep} objects as part of their required data.

On the other hand, the \class{OutputQuantity} class enables retrieval of specialized data during the course of simulations that are not normally found in the output files described above. Examples are force reactions pertaining to a particular portion of the external boundary, and results of special procedures such as $J$-integral\cite{Rice1968} calculations. In addition, derived classes may be used to implement integration of particular quantities over the problem domain (e.g.\ in order to calculate global error norms), or to set up ``observation points'' to track the evolution of specific quantities at a user-selected location.
\section{Vector and matrix operations} \label{sec:linearAlgebra}
For numerical schemes such as finite and boundary elements, a substantial portion of the total running time for a given simulation is spent on dense matrix and vector operations. It is therefore important that such calculations are performed efficiently, and in {\broomstyx} this is be done by making use of specialized BLAS implementations such as MKL\cite{IntelMKL} or OpenBLAS\cite{OpenBLAS}. Unfortunately, the interfaces provided by BLAS can be rather intimidating for those not well-versed in their usage. Consider for example the BLAS function \texttt{dgemv} that is used for matrix-vector multiplication with double floating point entries. It performs the operation $\mathbf{y} := \alpha \mathrm{op} \left( \mathbf{A} \right) \mathbf{x} + \beta \mathbf{y}$ for $\mathrm{op} \left( \mathbf{A} \right) = \mathbf{A}$ or $\mathbf{A}^\trp$ and  has the following declaration:
\begin{lstlisting}
void cblas_dgemv(const  CBLAS_LAYOUT Layout, const  CBLAS_TRANSPOSE TransA, const MKL_INT M, const MKL_INT N, const double alpha, const double *A, const MKL_INT lda, const double *X, const MKL_INT incX, const double beta, double *Y, const MKL_INT incY);
\end{lstlisting}
In order to provide a more user-friendly environment for developers, {\broomstyx} provides \class{RealVector} and \class{RealMatrix} classes as abstractions for dense vectors and matrices of real (type \texttt{double}) numbers, the latter storing their components in column-major format. Accordingly, the +,-,* and / operators are overloaded to accept matrix and vector arguments. They function as high-level wrappers that call appropriate BLAS routines to perform the actual matrix operations. In addition, they distinguish between \emph{lvalue} and \emph{rvalue} operands in order to minimize the occurrence of deep copies associated with temporary objects. The main purpose is to allow the coding of sequences of matrix and vector operations in a way that mimics standard mathematical notation but at the same time results in efficient evaluation. For instance given scalars $\alpha$ and $\beta$, matrix $\mathbf{A}$ of size $n\times m$ and vectors $\mathbf{b}$ and $\mathbf{c}$ of length $n$ and $m$ respectively, the expression
\begin{linenomath}
\begin{equation}
	\mathbf{d} = \alpha \mathbf{A}^\trp \mathbf{b} + \beta \mathbf{c}
	\label{eq:sampleMatVecOp}
\end{equation}
\end{linenomath}
can be coded as shown in Listing \ref{lst:syntaxMatrixOps}, with the back-end evaluation making use of the BLAS functions \texttt{dscal}, \texttt{dgemv} and \texttt{daxpy}.
\begin{lstlisting}[float, caption={User-friendly syntax for coding matrix operations.},
label={lst:syntaxMatrixOps}]
int m, n;
... 
double alpha, beta;
RealMatrix A(n,m);
RealVector b(n), c(m), d;
...
d = alpha*trp(A)*b + beta*c;
\end{lstlisting}
As operator overloading involves static polymorphism, no run-time overhead is incurred due to virtual function calls. Nevertheless, a naive implementation of the operator overloading results in an inefficient calculation as the binary nature of the aforementioned operators prevents us from taking full advantage of the low level BLAS functions' capabilities. Such implementation would yield for instance the sequence of operations shown in Table \ref{tab:naiveImplementation}.
\begin{table}
	\centering
	\caption{Sequence of operations for $\mathbf{d} = \alpha \mathbf{A}^\trp \mathbf{b} + \beta \mathbf{c}$ using naive operator overloading.}
	\label{tab:naiveImplementation}
	\begin{tabular}{ll}
		\toprule 
		Operation & Performed task / low-level BLAS call \\
		\toprule 
		$\mathbf{T}_1 \leftarrow \mathbf{A}^\trp$ &  heap allocation + manual transposition \\
		$\mathbf{T}_1 \leftarrow \alpha\mathbf{T}_1$ & \texttt{dscal} \\
		$\mathbf{t}_2 \leftarrow \mathbf{T}_1 \mathbf{b}$ & heap allocation + \texttt{dgemv} \\
		$\mathbf{t}_3 \leftarrow \beta\mathbf{c}$ & heap allocation + \texttt{daxpy} \\
		$\mathbf{t}_2 \leftarrow \mathbf{t}_2 + \mathbf{t}_3$ & \texttt{daxpy} \\
		$\mathbf{d} \leftarrow \mathbf{t}_2$ & move assignment \\
		\bottomrule
	\end{tabular}
\end{table}
These operations must preserve the original input arguments $\alpha$, $\beta$, $\mathbf{A}$, $\mathbf{b}$ and $\mathbf{c}$. Accounting for the move assignment in the final step, we see that two intermediate objects are created which are discarded at the end of the sequence, namely $\mathbf{T}_1$ and $\mathbf{t}_3$. Furthermore, manual transposition is performed and four calls to the low-level BLAS routines are made. In contrast, Table \ref{tab:optimizedImplementation} shows an optimized sequence consisting of only two operations that take full advantage of BLAS capabilities. In particular, no transposition of the matrix components needs to be done before hand since this can be handled internally by \texttt{dgemv}. Also, no intermediate objects are created.
\begin{table}
	\centering
	\caption{Optimized sequence of operations for $\mathbf{d} = \alpha \mathbf{A}^\trp \mathbf{b} + \beta \mathbf{c}$.}
	\label{tab:optimizedImplementation}
	\begin{tabular}{ll}
		\toprule 
		Operation & Performed task / low-level BLAS call \\
		\toprule 
		$\mathbf{d} \leftarrow \alpha \mathbf{A}^\trp \mathbf{b}$ & heap allocation + \texttt{dgemv} \\
		$\mathbf{d} \leftarrow \mathbf{d} + \beta\mathbf{c}$ & \texttt{daxpy} \\
		\bottomrule
	\end{tabular}
\end{table}
In order to approach the efficiency of using direct BLAS calls but at the same time keep the user-friendly syntax shown in Listing \ref{lst:syntaxMatrixOps}, we employ a lazy evaluation strategy when it comes to matrix transposition and scalar multiplication operations. A similar strategy has been adopted previously in the C++ library FLENS\cite{Lehn2005}, which uses expression templates to implement an approach based on the concept of \emph{closures} from functional programming. In the current work, this lazy evaluation strategy is achieved by designing the \class{RealVector} and \class{RealMatrix} classes to store a scaling factor as well as the method \method{isScaled()} that returns a value of \texttt{false} if the scaling factor is 1.0 and \texttt{true} otherwise. In addition, the latter class also stores a transposition flag and an associated \method{isTransposed()} for accessing its value. Finally, both classes provide the method \method{ownsItsPointer()} that returns a boolean value indicating whether a particular instance owns the memory address space referenced by its pointer, or if said memory belongs to another vector or matrix object. Consequently, the operator overload implementation for the scalar-matrix multiplication operation $\alpha \mathbf{A}$ needs only to return a matrix object whose scaling factor is multiplied by $\alpha$ and whose internal data pointer references the memory address in which the components of $\mathbf{A}$ are stored. The same trick is exploited in implementing the matrix transpose, so that for example the operation \texttt{B = trp(trp(A))} does not actually carry out any transposition. The state in which an object is transposed, has non-unity scaling or non-ownership of its pointer is only allowed to occur for \emph{r-value} objects, and to enforce this the copy and move assignment operators call a method \method{simplify()} that performs the following tasks: a) for matrices, perform manual transposition if the current object is in a transposed state, b) perform a deep copy of the matrix/vector components if the current object does not own its pointer, and c) scale the components accordingly if the scaling factor is not equal to 1, via the BLAS function \texttt{dscal}. In the case of matrices, the transposition is performed first as its implementation implies a deep copy (i.e., the actual transposition is not done ``in place''). With the aforementioned strategy, the execution of \eqref{eq:sampleMatVecOp} occurs according to the sequence shown in Table \ref{tab:broomstyxImplementation}.
\begin{table}
	\centering
	\caption{Sequence of operations for $\mathbf{d} = \alpha \mathbf{A}^\trp \mathbf{b} + \beta \mathbf{c}$ as implemented in {\broomstyx}.}
	\label{tab:broomstyxImplementation}
	\begin{tabular}{ll}
		\toprule 
		Operation & Performed task / low-level BLAS call \\
		\toprule 
		$\mathbf{T}_1^{\left( \mathbf{A}, 1, \trp \right)} \leftarrow \mathbf{A}^\trp$ &  set transpose flag and pointer \\
		$\mathbf{T}_1^{\left( \mathbf{A}, \alpha, \trp \right)} \leftarrow \alpha\mathbf{T}_1^{\left( \mathbf{A}, 1, \trp \right)}$ & set scaling factor \\
		$\mathbf{t}_2 \leftarrow \mathbf{T}_1^{\left( \mathbf{A}, \alpha, \trp \right)} \mathbf{b}$ & heap allocation + \texttt{dgemv} \\
		$\mathbf{t}_3^{\left( \mathbf{c}, \beta \right)} \leftarrow \beta\mathbf{c}$ & set pointer and scaling factor \\
		$\mathbf{t}_2 \leftarrow \mathbf{t}_2 + \mathbf{t}_3^{\left( \mathbf{c}, \beta \right)}$ & \texttt{daxpy} \\
		$\mathbf{d} \leftarrow \mathbf{t}_2$ & move assignment \\
		\bottomrule
	\end{tabular}
\end{table}
Similar to the earlier naive procedure, we can see that the same number of intermediate objects are created, i.e., $\mathbf{T}_1$ and $\mathbf{t}_3$. However in the current sequence there is no memory allocation done for each of these objects; instead the pointer members of $\mathbf{T}_1$ and $\mathbf{t}_3$ reference memory belonging to $\mathbf{A}$ and $\mathbf{b}$ respectively. Furthermore, the BLAS calls are now identical to those appearing in Table \ref{tab:optimizedImplementation}. Since the intermediate objects are created by functions implementing the operator overloads (as opposed to within a class instance located on the heap), the objects themselves reside in stack memory and thus very little overhead is incurred with their creation. For the example discussed above, return value optimization (RVO) by the compiler reduces the overhead to one call each of the \class{RealMatrix} and \class{RealVector} move constructors for the procedure of Table \ref{tab:broomstyxImplementation}. Thus in order for the operations to be lumped together as desired, the move constructor should not perform simplification. This introduces an unwanted side effect, as illustrated in the following section of code:
\begin{lstlisting}
RealMatrix A, B, C;
...
RealMatrix D = 0.5*trp(A + B*C); // D is move-constructed
D(0,0) = 4.5;                    // will throw exception

RealMatrix E;
E = 0.5*trp(A + B*C); // Move assignment calls .simplify()
E(0,0) = 4.5;         // will not throw exception
\end{lstlisting}
In line 3, the right hand side is the result of an operation and thus matrix \texttt{D} would normally be instantiated via move construction. However as the latter does not invoke the \method{simplify()} method, at line 4 the matrix \texttt{D} is in a transposed state and has a scaling factor of 0.5 that has not yet been multiplied to its stored components. Furthermore, the C++11 standard allows the compiler to perform a form of copy elision known as named return value optimization (NRVO), which omits the call to copy/move constructors. This makes the access of \texttt{D(i,j)} problematic: while one can easily deal with a transposed state by returning a reference based on row-major access, the latter cannot be made to carry additional information regarding the scaling factor.\footnote{In the specific example given, the line \texttt{D(0,0) = 4.5} would require a value of 9 (and not 4.5) to be written into the appropriate place in memory, since the stored value should later simplify to 4.5 after the scaling is applied.} Moreover if the object does not own the memory referenced by its pointer, then performing a write operation has a the unintended outcome of also modifying the original referenced object. In order to prevent such occurrences, access of matrix and vectors components via the \texttt{()} operator is locked while the objects are in an unsimplified state, and the operator throws an exception when called unless both the transpose and scaling flag are set to \texttt{false}, and the object in question owns the memory referenced by its pointer. The aforementioned scenario can be avoided by simply rewriting the sequence of instructions, as demonstrated in lines 6--8 of the above code snippet.
\section{Software Parallelization} \label{sec:parallelization}
The current implementation of {\broomstyx} supports shared-memory parallelism, which is achieved through the use of OpenMP directives. As the \class{RealMatrix} and \class{RealVector} classes store their components as contiguous arrays with unit stride, performing operations on these objects in parallel is likely to result in \emph{false sharing} whereby different processors attempt to modify data residing on the same cache line, leading to forced local cache updates. The execution model adopted in {\broomstyx} is such that all local matrix and vector contributions associated with a computational cell are handled by a single thread. An added advantage of this is that developers implementing new numerics and materials in the framework essentially write serial code without having to worry about particular details regarding parallelization.

It is common for OpenMP-accelerated software to have code regions that run serially interspersed between blocks that are executed in parallel. A common example of this is when the software has to perform I/O operations, although this can also be made to run such that a single thread executes write operations to disk while the remaining threads are already running calculations pertaining to the subsequent time step. While such procedure has not yet been implemented in {\broomstyx}, we consider this a minor drawback since for heavily nonlinear problems that require a large number of iterations, the time spent doing computation far outweighs the amount of time it takes to perform serial writing of output. Nevertheless, it is beneficial to reduce the amount of time in which the software is running serial code in order to fully benefit from the multi-core architecture of present day CPUs. To this end, assembly of global matrices and vectors is done in parallel, and makes use of atomic operations to avoid data race conditions. On the other hand, solution of the resulting linear system is done via calls to third-party solvers. Consider for example the direct solver PARDISO, which implements a solution of the linear system based on a sparse LU decomposition of the global coefficient matrix. An advantage of direct solvers is that one does not need to provide preconditioners that are often necessary to accelerate iterative schemes, and which may be difficult to construct for general nonlinear systems. Their drawback is that the cost of the LU decomposition becomes prohibitive for very large problem sizes. The PARDISO interface allows for splitting of the solution procedure into four phases:
\begin{enumerate}[a)]
	\setlength{\itemsep}{0pt}
	\setlength{\parskip}{0pt}
	\setlength{\parsep}{0pt}
	\item Reordering, symbolic factorization and allocation of memory to be used by the solver
	\item numerical factorization
	\item forward and back-substitution
	\item Release of allocated memory
\end{enumerate}
The first step is the most costly, taking up to 75\% of the total solve time and is also not easily executed in parallel. Many problems of interest however can be solved without performing adaptive refinement of the initial spatial discretization, which means that the sparsity profile of the global coefficient matrix does not change over time. Nonlinear solution algorithms implemented in {\broomstyx} exploit this property by performing symbolic factorization only at the beginning iteration of a substep, and then skipping this step in subsequent iterations along with the release of memory unless the solver returns an error, in which case the stalled solution is restarted going through all the phases listed above. We have found that when low order methods are employed for the discrete problem, use of the above strategy can reduce the total time for each iteration (i.e., assembly + solve) by as much as 50\%.

\section{Example applications} \label{sec:examples}

In this section, we present several numerical examples dealing on different topics and that are specifically chosen to highlights important capabilities of the {\broomstyx} framework. We note that while all presented examples are in 2D, the code structure itself readily admits formulations in both lower and higher dimensions. The first example deals with torsion and shows the code's capability to deal with higher order mappings and formulations as well as non-standard degrees of freedom that are not associated with specific geometric entities. The next two examples feature a monolithic coupling of variational and control volume formulations applied to the problem of poroelasticity. The final example deals with brittle fracture propagation in concrete using the phase-field approach, which gives rise to a coupled nonlinear system of PDEs that must be solved alternately with a Newton approach applied to the inner iterations. For all problems, discretization of the problem domain was performed using the Gmsh software\cite{Geuzaine2009}, while 2D color plots were generated either Paraview and Gmsh, the latter used for higher order visualization.

\subsection{Elastic torsion of composite shafts} \label{sec:StVenantTorsion}
The Saint-Venant torsion problem is concerned with determining the shear stresses induced in a non-circular prismatic shaft that is rigidly clamped at one end and subjected to an applied torque at the other. This problem has been extensively studied in the literature, and a discussion of solution approaches can be found in classic texts such as Fung\cite{Fung1965}. Numerical solutions are generally designed based on Saint-Venant's original approach of using a warping function, or alternatively on the Prandtl stress function. Both lead to Laplace/Poisson type governing equations but with notable differences in the resulting boundary conditions and constraints. In this example we make use of the former strategy, its chief advantage being that no special treatment is needed when dealing with multiply connected sections.

We begin by considering a prismatic shaft in Cartesian space, whose longitudinal axis is aligned with the $z$-direction and whose cross section normal to said axis is denoted by $\Omega$. If $\Omega$ is non-circular then it will experience warping when the shaft is twisted, however it can be assumed that its projection onto the $\left( x,y \right)$-plane rotates as a rigid body. For a general section, the displacement at any point $\mathbf{x}$ may be expressed as follows\cite{Weinstein1947}:
\begin{linenomath}
\begin{equation}
	\mathbf{u} = \beta \left\{ \begin{array}{ccccc}
	-zy & + & a_x & + & k_y z - k_z y \\
	zx & + & a_y & + & k_z x - k_x z \\
	\omega \left( x,y \right) & + & a_z & + & k_x y - k_y x
	\end{array} \right\},
\end{equation}
\end{linenomath}
wherein $\beta \ll 1$ is the angle of twist per unit length, $\omega \left( x,y \right)$ is the so-called warping function, and $\mathbf{a} + \mathbf{k} \times \mathbf{x}$ is a rigid body motion consisting of a pure translation component plus pure rotation that are both a priori unknown. The rigid clamping at $z = 0$ corresponds to the conditions $u_x \left( x,y,0 \right) = u_y \left( x,y,0 \right) = u_z \left( x,y,0 \right) = 0$, and imposition of the first two yields
\begin{linenomath}
\begin{equation}
	a_x = a_y = k_z = 0,
\end{equation}
\end{linenomath}
so that the displacement simplifies to
\begin{linenomath}
	\begin{subequations}
		\begin{align}
		u_x &= -\beta z \left( y - k_y \right) \\
		u_y &= \beta z \left( x - k_x \right) \\
		u_z &= \beta \left[ \omega \left( x,y \right) + a_z + k_x y - k_y x \right]
		\label{eq:stVenantZDisp}
		\end{align}
		\label{eq:stVenantDisplacement}
	\end{subequations}
\end{linenomath}
On the other hand, the constraint $u_z \left( x,y,0 \right)$ cannot be directly imposed; instead it is weakly enforced via the condition
\begin{linenomath}
\begin{equation}
	\int_\Omega \left( u_z \right)^2 \dee\Omega = \text{minimum}
	\label{eq:stVenantConstraint}
\end{equation}
\end{linenomath}
One can see from \eqref{eq:stVenantDisplacement} that $u_x$ and $u_y$ vanish when $x = k_x$ and $y = k_y$, hence the location $\left( k_x, k_y \right)$ is termed the center of twist for the cross section. For simplicity, we assume that $\beta$ is prescribed so that the relevant unknowns are the function $\omega \left( x,y \right)$ and the quantities $a_z$, $k_x$ and $k_y$. Furthermore, we assume that the resulting strains are infinitesimal, i.e.,
\begin{linenomath}
\begin{equation}
	\epsilon_{ij} = \frac{1}{2} \left( \frac{\partial u_j}{\partial x_i} + \frac{\partial u_i}{\partial x_j} \right), \qquad \gamma_{ij} = 2\epsilon_{ij}.
\end{equation}
\end{linenomath}
Plugging in the displacement expressions into the above formula, we find that only two strain components are nonzero, these being $\gamma_{zx}$ and $\gamma_{zy}$. If the component materials of the shaft are both linear elastic and isotropic, the corresponding stresses are given by
\begin{linenomath}
\begin{equation}
	\sigma_{zx} = G \beta \left( \frac{\partial\omega}{\partial x} - y \right) \qquad \sigma_{zy} = G \beta \left( \frac{\partial\omega}{\partial y} + x \right)
	\label{eq:stVenantStresses}
\end{equation}
\end{linenomath}
with $\sigma_{xx} = \sigma_{yy} = \sigma_{zz} = \sigma_{xy} = 0$. Consequently, the stress equilibrium equation in the absence of body forces reduces to the Laplace equation,
\begin{linenomath}
\begin{equation}
	\nabla^2 \omega = 0 \text{ in } \Omega.
		\label{eq:stVenantStrongForm}
\end{equation}
\end{linenomath}
For inhomogeneous shafts, the surface traction $\mathbf{t} = \mathbf{\sigma} \cdot \mathbf{n}$ must be continuous across the material interfaces. Equation \eqref{eq:stVenantStresses} can be written in a more compact form as $\mathbf{\sigma} = G\beta \left( \nabla\omega - \mathbf{x}_\perp \right)$ where $\mathbf{x}_\perp = \left\{ y, -x \right\}^\mathrm{T}$. Consequently, the previous condition can be expressed as
\begin{linenomath}
\begin{equation}
	G^+ \left( \nabla\omega^+ - \mathbf{x}_\perp \right) \cdot \mathbf{n} = G^- \left( \nabla\omega^- - \mathbf{x}_\perp \right) \cdot \mathbf{n} \implies G^+ \frac{\partial\omega}{\partial n}^+ - G^- \frac{\partial\omega}{\partial n}^- = \left( G^+ - G^- \right) \mathbf{x}_\perp \cdot \mathbf{n} \text{ on } \Gamma
	\label{eq:stVenantInternalBoundary}
\end{equation}
\end{linenomath}
where $\Gamma$ is an internal boundary separating two materials, and $\left( \bullet \right)^+$ and $\left( \bullet^- \right)$ denote quantities on either side of $\Gamma$. On the other hand for surfaces that are traction-free, the relevant condition is
\begin{linenomath}
	\begin{equation}
	\frac{\partial\omega}{\partial n} = \mathbf{x}_\perp \cdot \mathbf{n} \text{ on } \partial\Omega .
	\label{eq:stVenantBC}
	\end{equation}
\end{linenomath}
Finally for a given rate of twist, the resulting internal torque can be calculated as
\begin{linenomath}
\begin{equation}
	T = \int_\Omega \left( x \sigma_{zy} - y \sigma_{zx} \right) \dee\Omega = \beta \int_\Omega G \left( \mathbf{x} \cdot \mathbf{x} - \nabla\omega \cdot \mathbf{x}_\perp \right) \dee\Omega .
	\label{eq:stVenantTorque}
\end{equation}
\end{linenomath}

The complete boundary value problem is given by equations \eqref{eq:stVenantStrongForm}, \eqref{eq:stVenantInternalBoundary}, \eqref{eq:stVenantBC} together with the constraint \eqref{eq:stVenantConstraint}. The weak form corresponding to \eqref{eq:stVenantStrongForm}--\eqref{eq:stVenantBC} is
\begin{linenomath}
\begin{equation}
	\int_\Omega \nabla\delta\omega \cdot \nabla\omega \,\dee\Omega = \int_{\partial\Omega} \delta\omega \mathbf{x}_\perp \cdot \mathbf{n} \,\dee\Gamma + \sum\limits_{\Gamma_i} \int_{\Gamma_i} \delta\omega \left( G^+ - G^- \right) \mathbf{x}_\perp \cdot\mathbf{n} \,\dee\Gamma
	\label{eq:stVenantWeakForm}
\end{equation}
\end{linenomath}
where $\Gamma_i$ are the material interfaces, and $\delta\Omega$ denotes all internal and external boundaries that are traction-free. A further three equations are obtained by carrying out the minimization in \eqref{eq:stVenantConstraint} with respect to $a_z$, $k_x$ and $k_y$. This yields
\begin{linenomath}
\begin{align}
	&\int_\Omega \left[ \omega \left( x,y \right) + a_z + k_x y - k_y x \right] \dee\Omega = 0 \label{eq:azConstraint} \\
	&\int_\Omega y \left[ \omega \left( x,y \right) + a_z + k_x y - k_y x \right] \dee\Omega = 0 \label{eq:kxConstraint} \\
	&\int_\Omega -x \left[ \omega \left( x,y \right) + a_z + k_x y - k_y x \right] \dee\Omega = 0 \label{eq:kyConstraint}
\end{align}
\end{linenomath}
To obtain a finite element discretization of the above problem, let $\omega$ and $\delta\omega$ and their derivatives be approximated by
\begin{linenomath}
\begin{equation}
\begin{split}
	\omega = \sum_{I=1}^m N_I \hat{\omega}_I  \qquad \delta\omega = \sum_{I=1}^m N_I \delta\hat{\omega}_I, \\
	\nabla\omega = \sum_{I=1}^m \mathbf{B}_I \hat{\omega}_I  \qquad \nabla\delta\omega = \sum_{I=1}^m \mathbf{B}_I \delta\hat{\omega}_I, \\
\end{split}
\end{equation}
\end{linenomath}
where $\hat{\omega}_I$ and $\delta\hat{\omega}_I$ are values of $\omega$ and $\delta\omega$ at node $I$, $N_I$ are standard Lagrange-type shape functions, $\mathbf{B}_I = \nabla N_I$, and $m$ is the number of nodes. Plugging the above approximations into \eqref{eq:stVenantWeakForm}--\eqref{eq:kyConstraint} which must then hold for arbitrary values of $\delta\hat{\omega}_I$, we obtain the following system of equations in matrix form:
\begin{linenomath}
\begin{equation}
	\left[ \begin{array}{ccccc}
	\mathbf{K} & \mathbf{0} & \mathbf{0} & \mathbf{0} \\[0.3em]
	\mathbf{Q}^\trp & A & S_x & -S_y \\[0.3em]
	\mathbf{Q}_x^\trp & S_x & I_{xx} & -I_{xy} \\[0.3em]
	\mathbf{Q}_y^\trp & -S_y & -I_{xy} & I_{yy}
	\end{array} \right]
	\left\{ \begin{array}{c}
	\hat{\mathbf{\omega}} \\[0.3em] a_z \\[0.3em] k_x \\[0.3em] k_y
	\end{array}	\right\} =
	\left\{ \begin{array}{c}
	\mathbf{F}^{\partial\Omega} + \sum_i \mathbf{F}^{\Gamma_i} \\[0.3em] 0 \\[0.3em] 0 \\[0.3em] 0
	\end{array}	\right\},
	\label{eq:stVenantDiscreteProblem}
\end{equation}
\end{linenomath}
where the individual entries are given by
\begin{linenomath}
\begin{equation}
	\begin{array}{lll}
	K_{IJ} = \displaystyle \int_\Omega G \mathbf{B}_I^\trp \mathbf{B}_J  \,\dee\Omega \hspace*{2em} & 
	F_I^{\partial\Omega} = \displaystyle \int_{\partial\Omega} G N_I \mathbf{x}_\perp \cdot \mathbf{n} \,\dee\Gamma \hspace*{2em} 
	& F_I^{\Gamma_i} = \displaystyle \int_{\Gamma_i} \left( G^+ - g^- \right) N_I \mathbf{x}_\perp \cdot \mathbf{n} \,\dee\Gamma \\[1.2em]
	Q_I = \displaystyle \int_\Omega N_I \,\dee\Omega 
	& Q_{xI} = \displaystyle \int_\Omega y N_I \,\dee\Omega 
	& Q_{yI} = \displaystyle \int_\Omega -x N_I \,\dee\Omega \\[1.2em]
	S_x = \displaystyle\int_\Omega y \,\dee\Omega 
	& S_y = \displaystyle \int_\Omega x \,\dee\Omega \\[1.2em]
	I_{xx} = \displaystyle \int_\Omega y^2 \,\dee\Omega 
	& I_{xy} = \displaystyle \int_\Omega xy \,\dee\Omega 
	& I_{yy} = \displaystyle \int_\Omega x^2 \,\dee\Omega .
	\end{array}
\end{equation}
\end{linenomath}
We note that \eqref{eq:stVenantDiscreteProblem} does not give a symmetric system. Furthermore, the subsystem consisting of the equation $\mathbf{K} \, \hat{\mathbf{\omega}} = \mathbf{F}^{\partial\Omega} + \sum_i \mathbf{F}^{\Gamma_i}$ cannot be decoupled from the global system as \eqref{eq:stVenantWeakForm} constitutes a pure Neumann problem and therefore does not have a unique solution in the absence of additional constraints.

The discrete problem described above is implemented in the class \class{StVenantTorsion\_Fe\_Tri6} which utilizes isoparametric 6-node ($P_2$) triangles. The coefficient matrix is assembled by looping over all the domain elements, while the right hand side terms are assembled by looping over 3-node 1D elements that have been marked as part of the boundary. Domain integrals are evaluated using a 3-point Gauss Legendre quadrature since the integrands are at most quadratic.
\footnote{This is not always true; in particular when an element is curved then the shape functions become rational due to the mapping between reference and physical spaces. However such a case only occurs for elements immediately adjacent to a curved boundary, and with sufficient refinement the error arising from under-integration can be minimized.}
Meanwhile, the boundary terms are calculated using a 3-point quadrature rule in 1D as explained in Appendix \ref{app:StVenantBCTerm}. From an implementation perspective, the system shown in \eqref{eq:stVenantDiscreteProblem} is interesting as the \class{Dof} objects corresponding to $a_z$, $k_x$ and $k_y$ are not connected to any particular element. Instead they are associated with the \class{Numerics} instance itself, which implies that only one such object should be instantiated for the entire problem; inhomogeneity is handled through the use of appropriate \class{Material} objects which provide the necessary shear modulus for each subdomain.

\begin{figure}
	\centering
	\begin{subfigure}{0.55\textwidth}
		\centering
		\includegraphics[width=\textwidth]{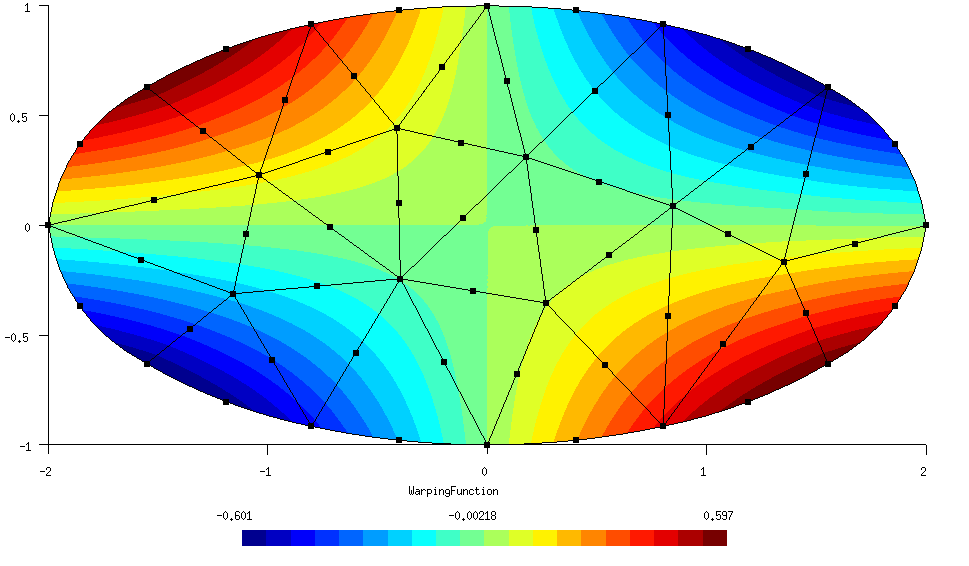}
		\caption{}
		\label{fig:warpFcnSoln}
	\end{subfigure}
	\begin{subfigure}{0.44\textwidth}
		\centering
		\includegraphics[width=\textwidth]{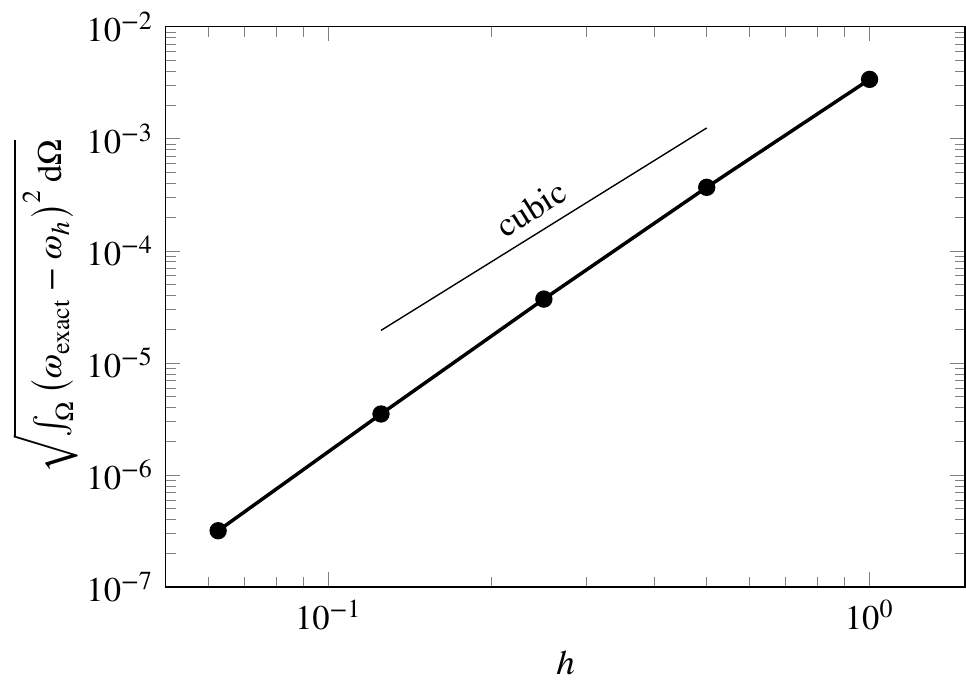}
		\caption{}
		\label{fig:ellipseConvergence}
	\end{subfigure}
	\caption{Saint-Venant torsion of a bar with elliptical cross section: (a) FE solution of the warping function $\omega$ using isoparametric 6-node triangles, and (b) convergence of $\omega$ to the analytical solution with hierarchical refinement of the mesh.}
\end{figure}

\begin{figure}
	\centering
	\begin{subfigure}{0.49\textwidth}
		\centering
		\includegraphics[width=\textwidth]{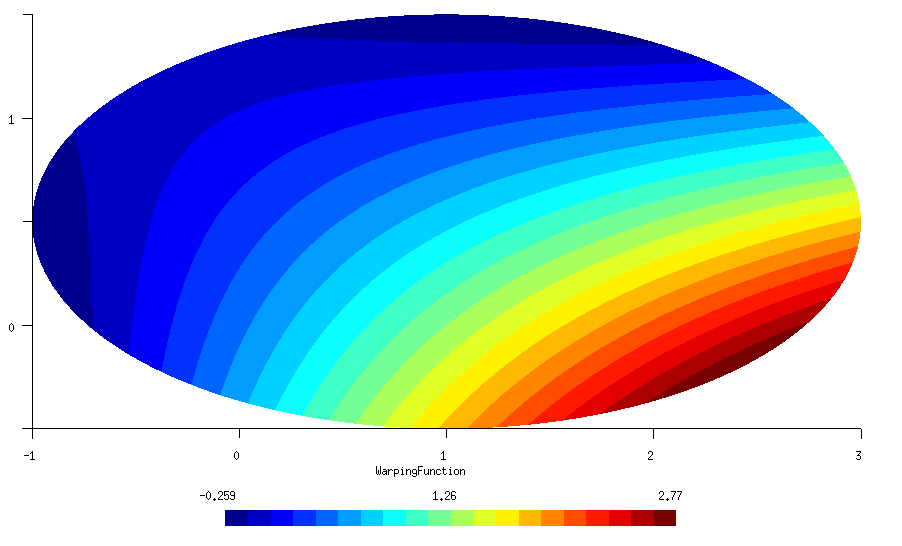}
		\caption{}
	\end{subfigure}
	\begin{subfigure}{0.49\textwidth}
		\centering
		\includegraphics[width=\textwidth]{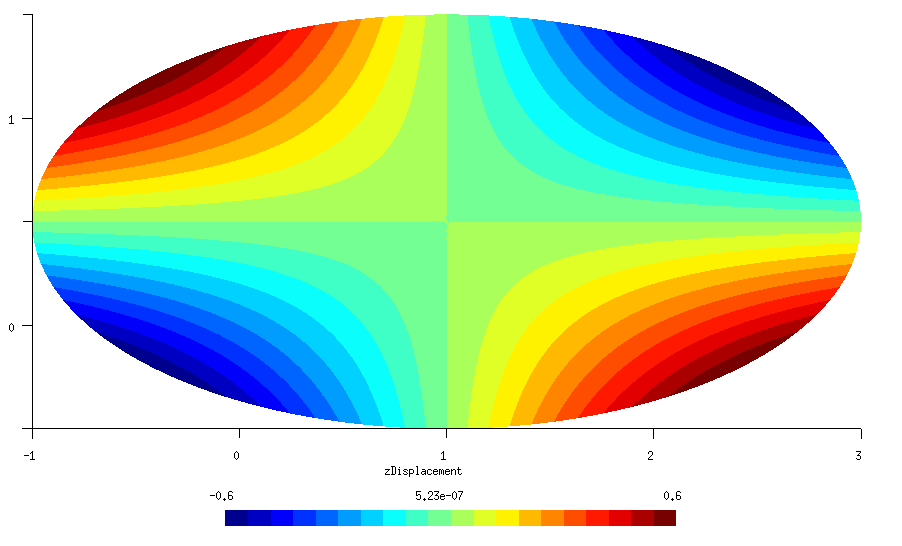}
		\caption{}
	\end{subfigure}
	\caption{Numerical results showing (a) the warping function $\omega$, and (b) the displacement component $u_z$ for an elliptical cross-section centered at $\left( 1,0.5 \right)$.}
	\label{fig:stVenantTranslatedEllipse}
\end{figure}

To verify the implementation, we examine the case of a shaft with an elliptical cross section centered at $\left( 0,0 \right)$ with major axis oriented along the $x$-direction. In this particular case the analytical solution for the warping function is given by\cite{Fung1965}
\begin{linenomath}
\begin{equation}
	\omega \left( x,y \right) = -\frac{a^2 - b^2}{a^2 + b^2}xy.
\end{equation}
\end{linenomath} 
where $a$ and $b$ are the respective lengths of the semi-major and semi-minor axes. For the numerical simulation, we take $a = 2$ and $b = 1$, and the shear modulus and rate of twist are also set to unity, i.e., $G = \beta = 1$. The finite element solution for $\omega \left( x,y \right)$ is plotted in Figure \ref{fig:warpFcnSoln} for the coarsest mesh used in the study.
Furthermore, we can see from Figure \ref{fig:ellipseConvergence} that the convergence rate with respect to mesh refinement is slightly better than cubic. This is to be expected for this particular case, i.e., even though the analytical solution is contained in the approximation space used, it is nonetheless impossible to reproduce it exactly as representation of the geometry with 6-node triangles results in the outer layer of elements having curved edges. Also, the warping function itself is physically meaningful only when the center of twist coincides with the origin. For cross sections having their center of twist elsewhere, the cross sectional warping is given by $u_z$ which must be calculated according to \eqref{eq:stVenantZDisp}. This is illustrated in Figure \ref{fig:stVenantTranslatedEllipse}.

\begin{figure}
	\centering
	\begin{subfigure}{0.32\textwidth}
		\centering
		\includegraphics[width=\textwidth]{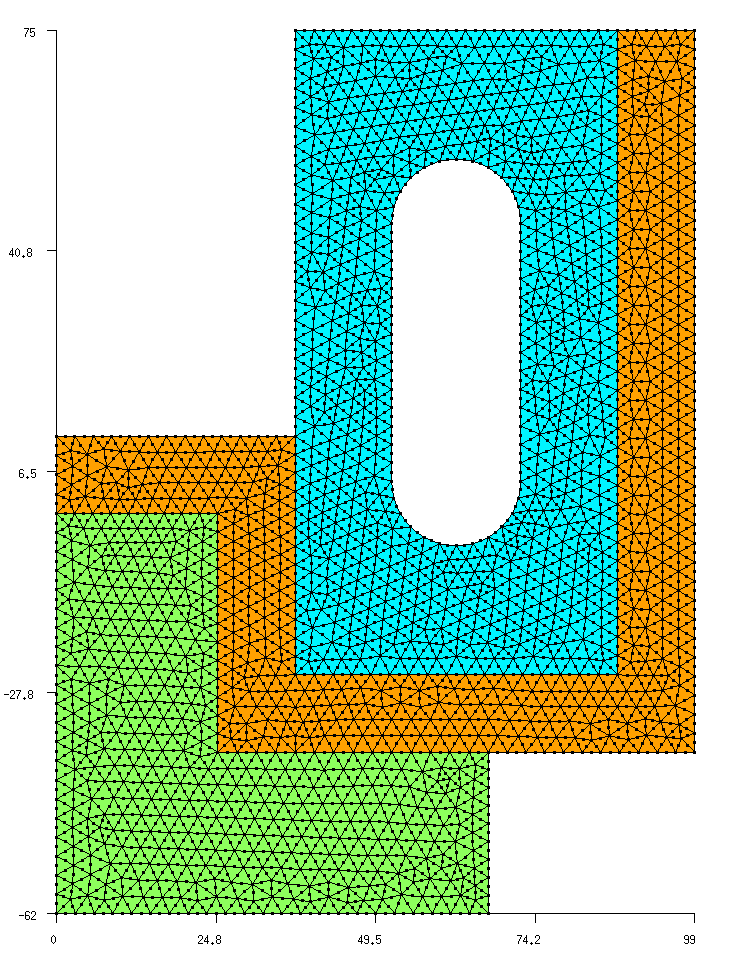}
		\caption{}
		\label{fig:compositeSectionMesh}
	\end{subfigure}
	\begin{subfigure}{0.32\textwidth}
		\centering
		\includegraphics[width=\textwidth]{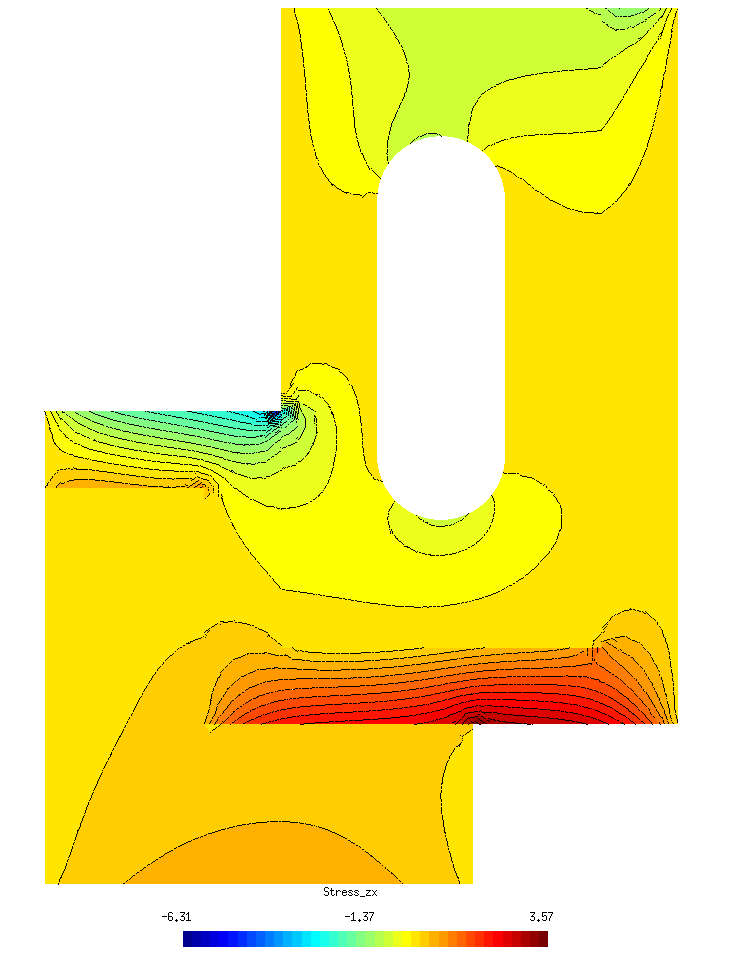}
		\caption{}
		\label{fig:stressContinuityZX}
	\end{subfigure}
	\begin{subfigure}{0.32\textwidth}
		\centering
		\includegraphics[width=\textwidth]{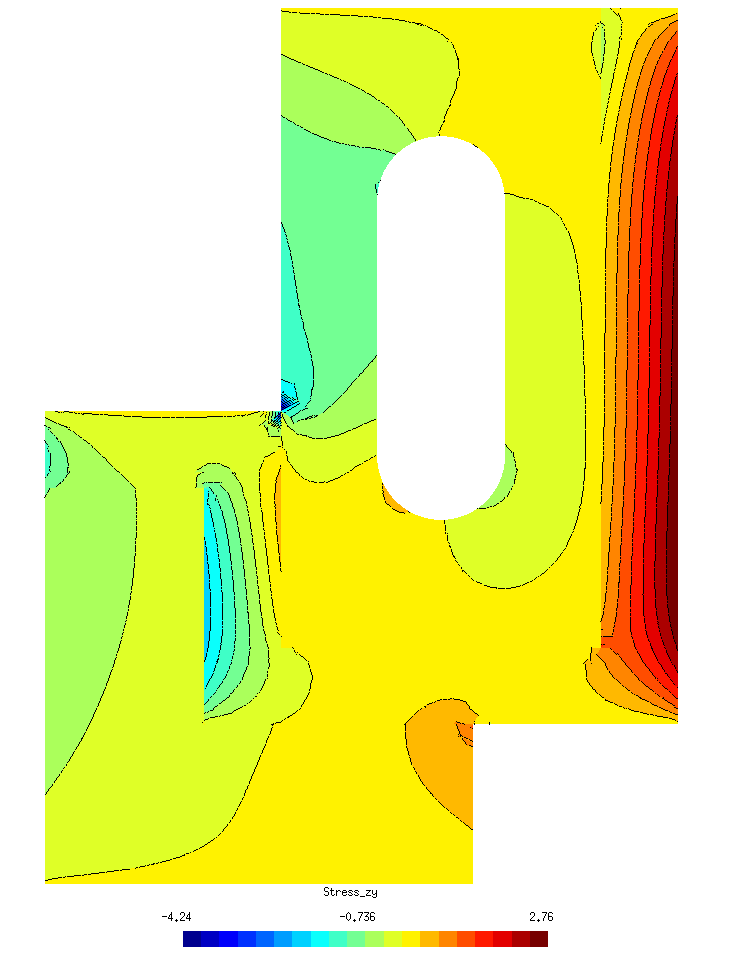}
		\caption{}
		\label{fig:stressContinuityZY}
	\end{subfigure}
	\caption{Torsion of composite shaft, showing (a) geometry and discretization, (b) distribution of $\sigma_{zx}$, (c) distribution of $\sigma_{zy}$. It can be verified from the latter two plots that continuity of traction components normal to material interfaces is properly enforced across said surfaces.}
\end{figure}

\begin{figure}
	\centering
	\hfill
	\begin{subfigure}{0.31\textwidth}
		\centering
		\includegraphics[width=\textwidth]{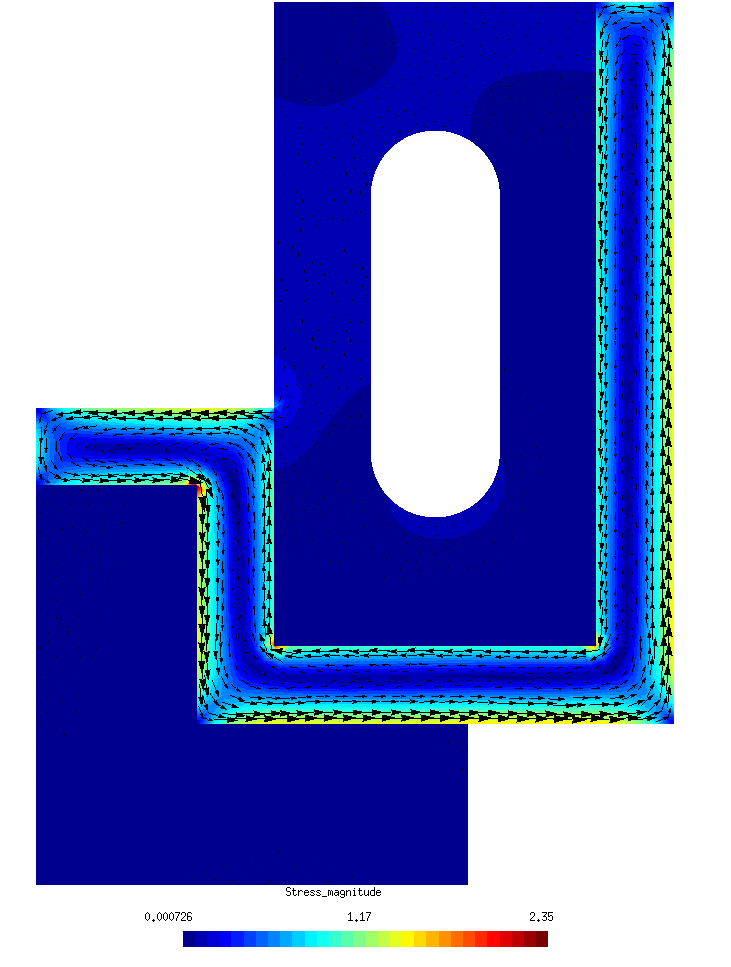}
		\captionsetup{skip=-0.2em}
		\caption{$G_A = 1$, $G_B=100$}
	\end{subfigure} \hfill
	\begin{subfigure}{0.31\textwidth}
		\centering
		\includegraphics[width=\textwidth]{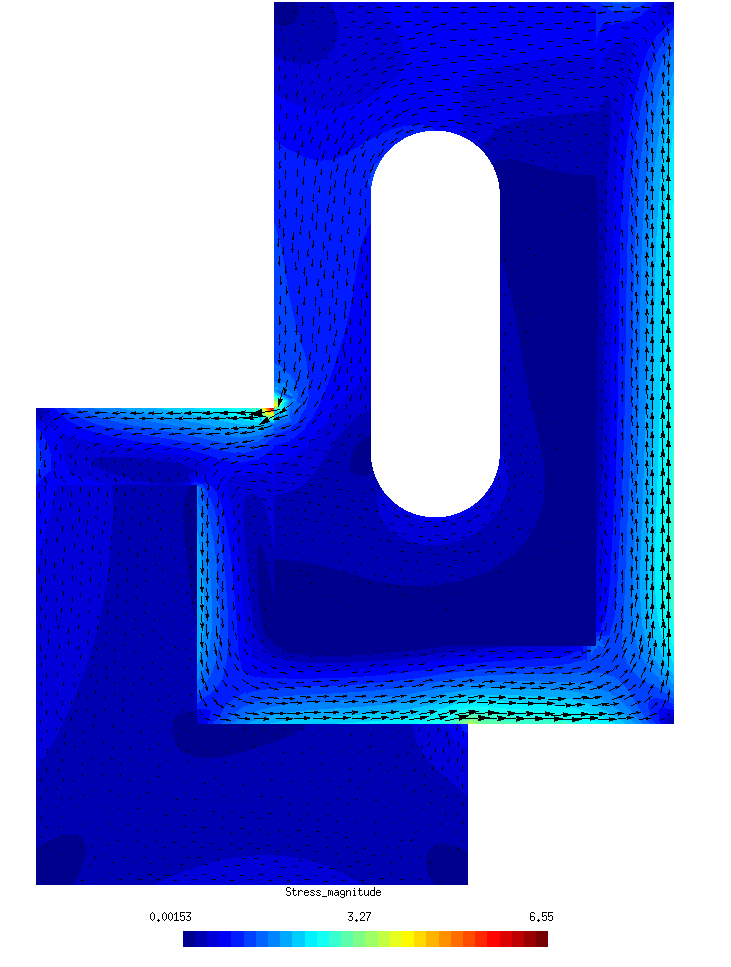}
		\captionsetup{skip=-0.2em}
		\caption{$G_A = 10$, $G_B=100$}
	\end{subfigure} \hfill
	\begin{subfigure}{0.31\textwidth}
		\centering
		\includegraphics[width=\textwidth]{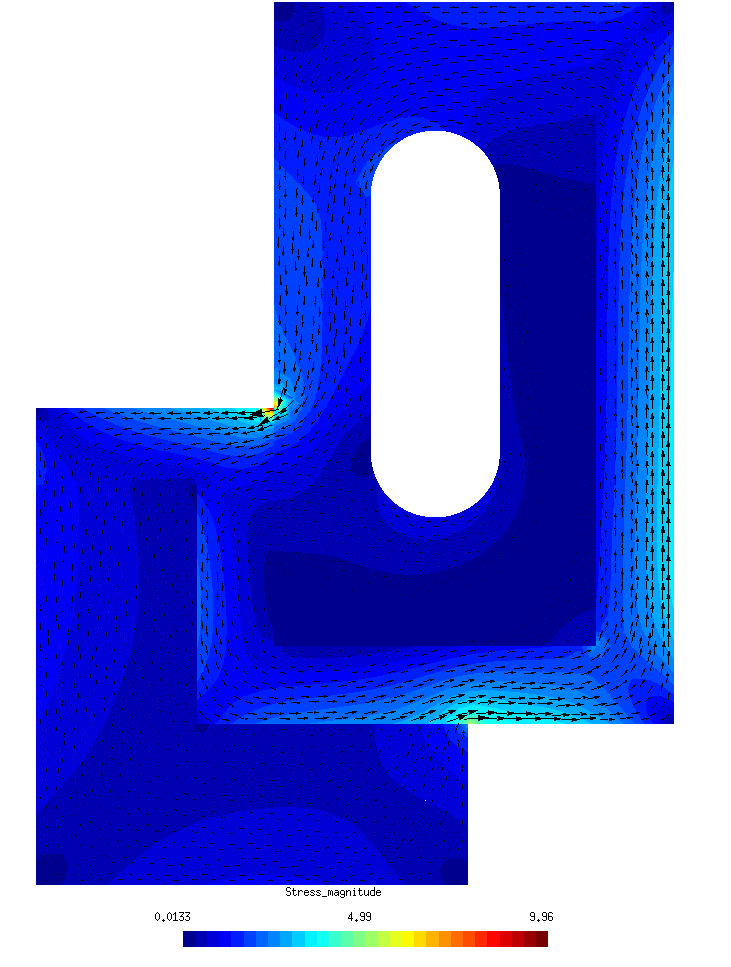}
		\captionsetup{skip=-0.2em}
		\caption{$G_A = 20$, $G_B=100$}
	\end{subfigure} \hfill \\[0.5em]
	\hfill
	\begin{subfigure}{0.31\textwidth}
		\centering
		\includegraphics[width=\textwidth]{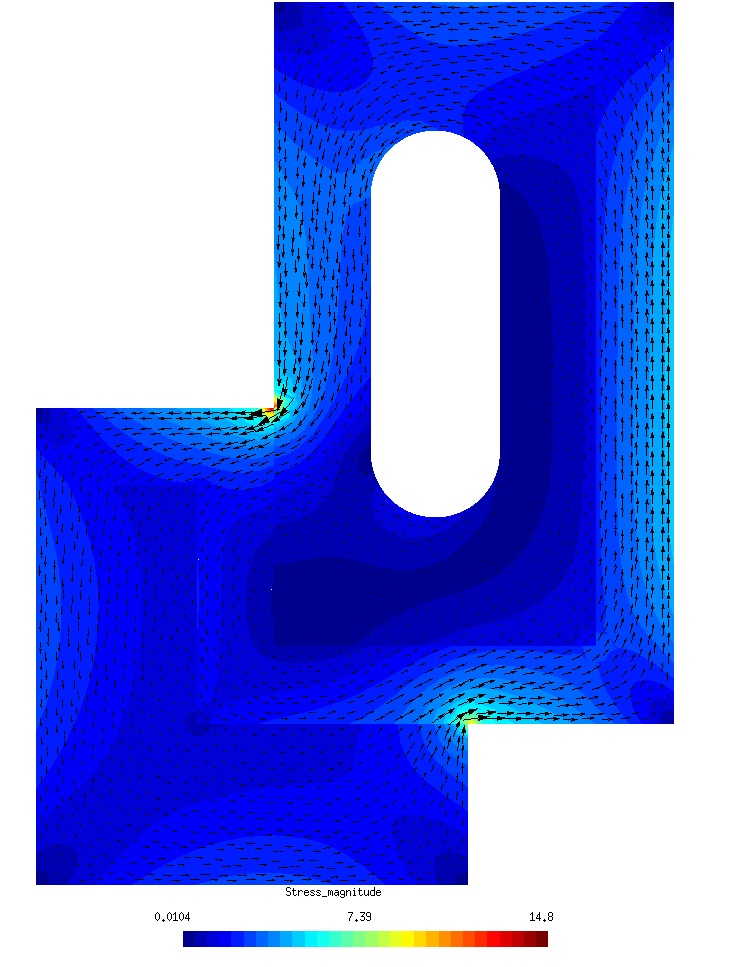}
		\captionsetup{skip=-0.2em}
		\caption{$G_A = 50$, $G_B=100$}
	\end{subfigure} \hfill
	\begin{subfigure}{0.31\textwidth}
		\centering
		\includegraphics[width=\textwidth]{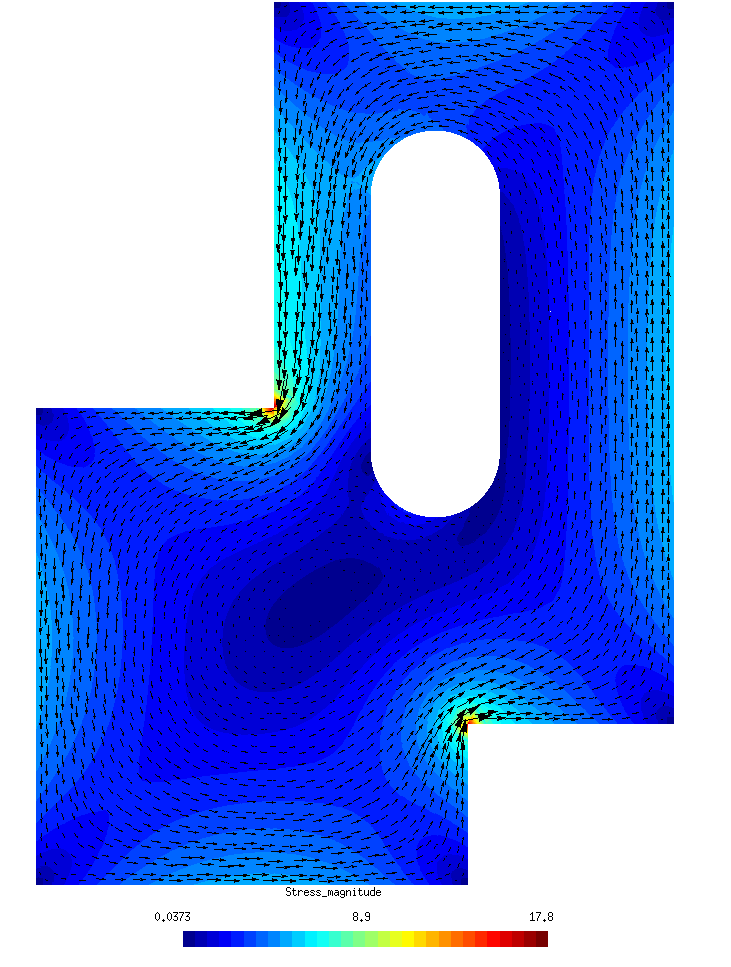}
		\captionsetup{skip=-0.2em}
		\caption{$G_A = 100$, $G_B=100$}
	\end{subfigure} \hfill
	\begin{subfigure}{0.31\textwidth}
		\centering
		\includegraphics[width=\textwidth]{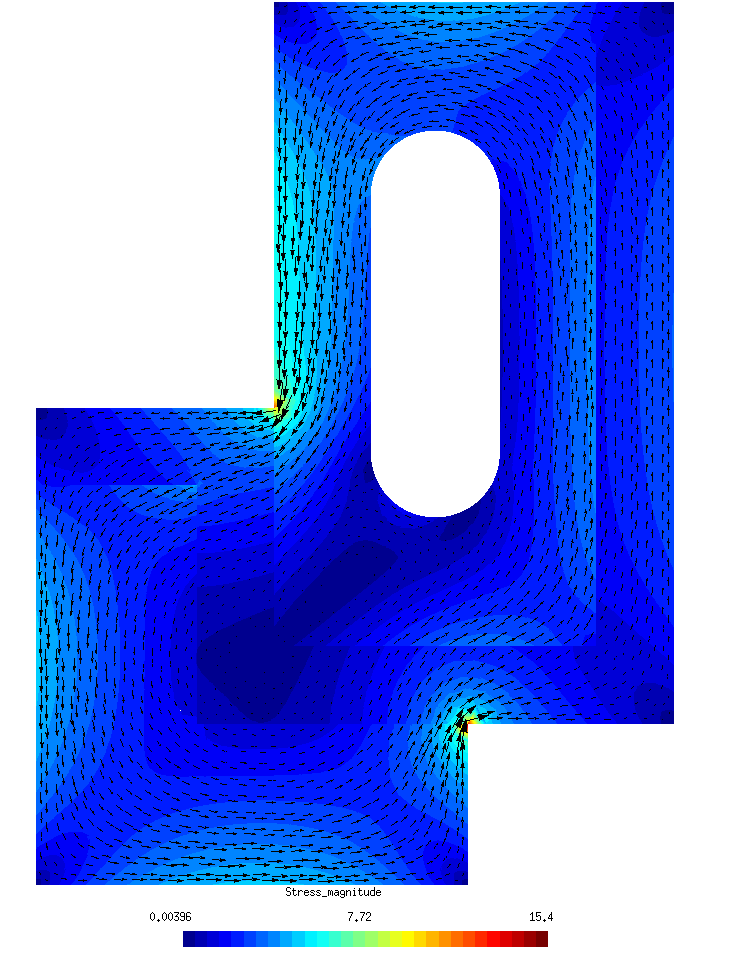}
		\captionsetup{skip=-0.2em}
		\caption{$G_A = 100$, $G_B=50$}
	\end{subfigure} \hfill \\[0.5em]
	\hfill
	\begin{subfigure}{0.31\textwidth}
		\centering
		\includegraphics[width=\textwidth]{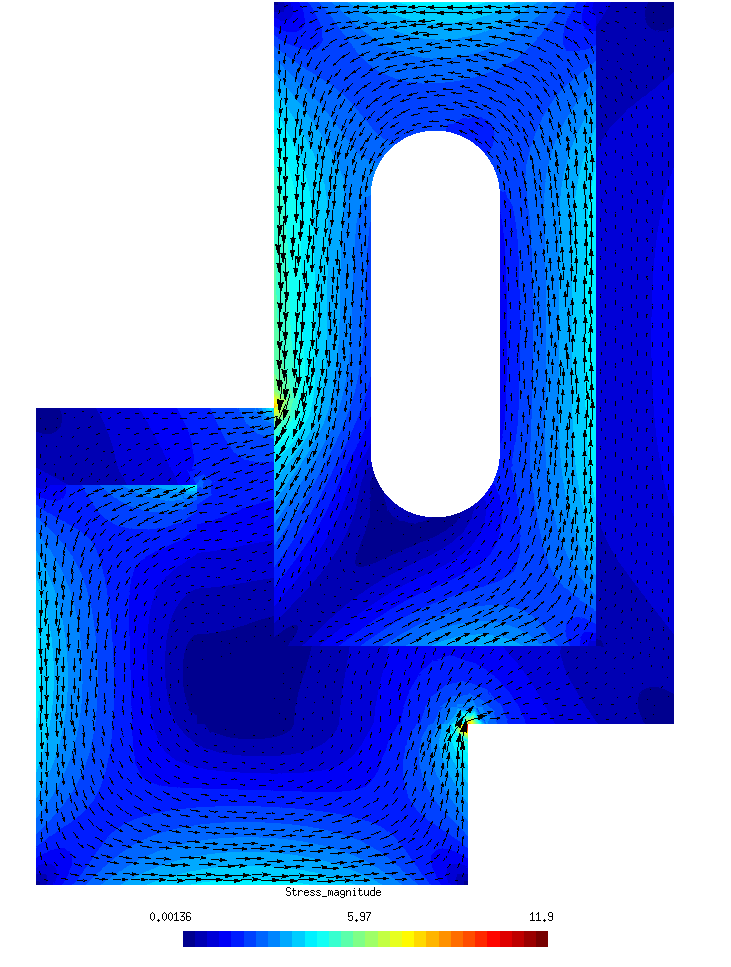}
		\captionsetup{skip=-0.2em}
		\caption{$G_A = 100$, $G_B=20$}
	\end{subfigure} \hfill
	\begin{subfigure}{0.31\textwidth}
		\centering
		\includegraphics[width=\textwidth]{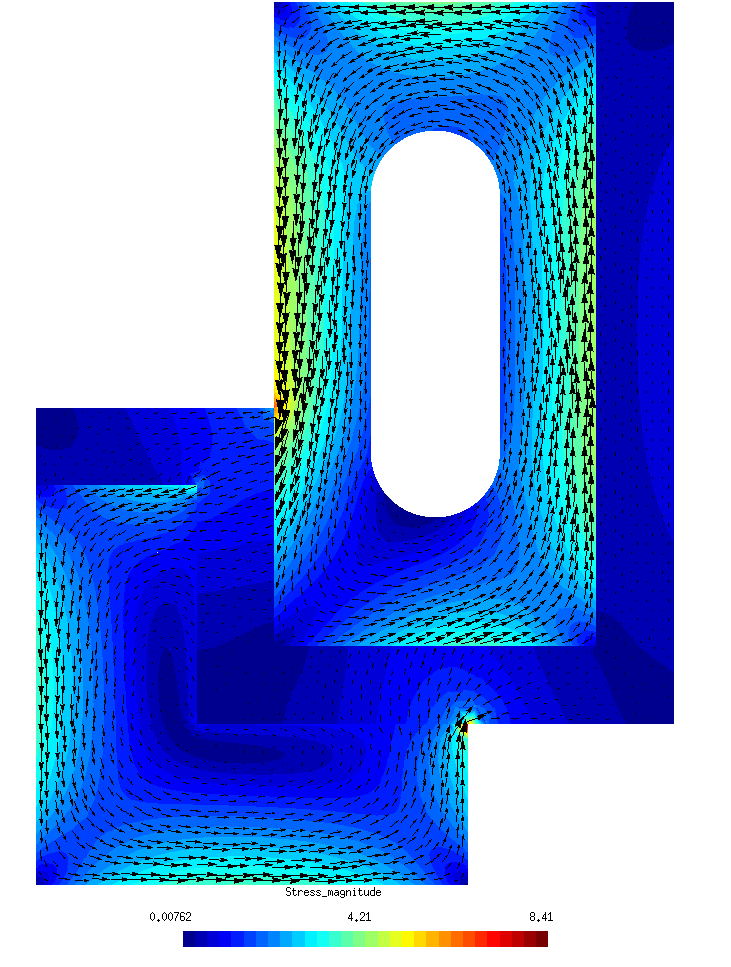}
		\captionsetup{skip=-0.2em}
		\caption{$G_A = 100$, $G_B=10$}
	\end{subfigure} \hfill
	\begin{subfigure}{0.31\textwidth}
		\centering
		\includegraphics[width=\textwidth]{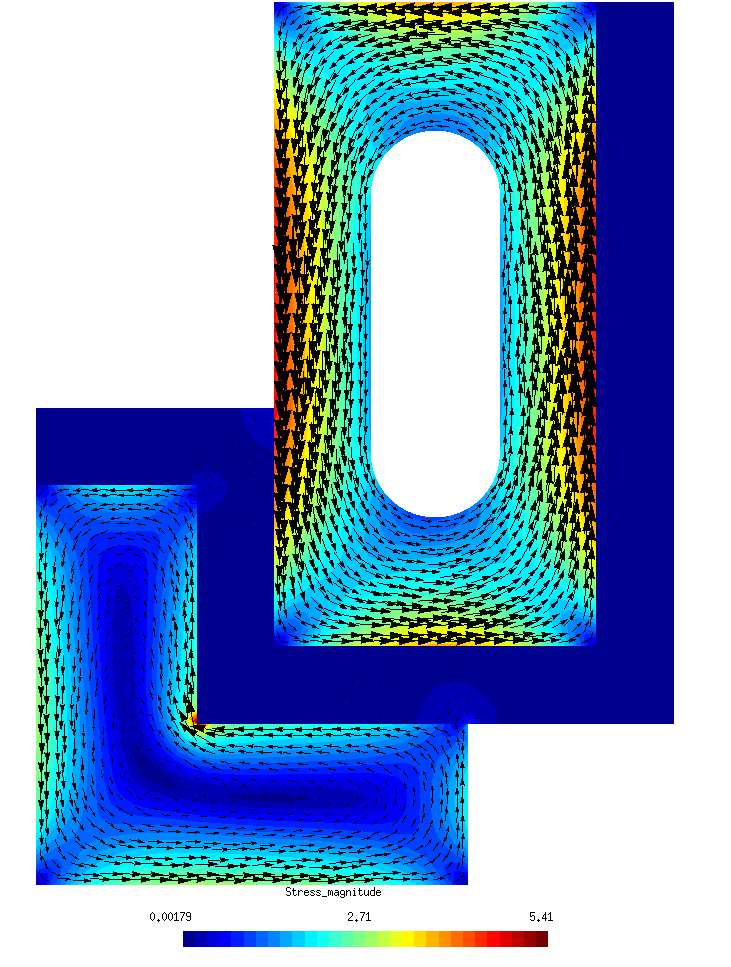}
		\captionsetup{skip=-0.2em}
		\caption{$G_A = 100$, $G_B=1$}
	\end{subfigure} \hfill
	\caption{Dependence of shear flow on ratio of shear modulus $G_A$ of the lower block and hollow insert to that of the middle section, $G_B$. Colors and arrows describe the magnitude and direction of the traction vector $\mathbf{t} = \left\{ \sigma_{zx}, \sigma_{zy} \right\}^\trp$.}
	\label{fig:ratioEffect}
\end{figure}

Next, we use the same \class{Numerics} class as above to analyze a composite shaft consisting of an L-shaped lower block, an angled middle section and a hollow insert as shown in Figure \ref{fig:compositeSectionMesh}.
Both the lower block and the hollow insert are made up of the same material, whose shear modulus is set to a value of $G_A = 10$, whereas the middle section has $G = G_B = 100$. The angle of twist $\beta$ is prescribed as 0.003. Once the proper mesh has been generated, setup of the problem is relatively straightforward as evident from Listing \ref{lst:compositeTorsion}.
\begin{lstlisting}[float,stringstyle=\color{black},caption={Input file for torsion of composite shaft.},label={lst:compositeTorsion}]
*MESH_READER GmshReader
*MESH_FILE CompositeSection.msh

*SOLUTION_STAGES 1
*FIELDS_PER_NODE 2
*FIELDS_PER_CELL 4

*DOF_PER_NODE 1
    1 DofGroup 1 NodalField 1 2
*DOF_PER_CELL 0

*NUMERICS 1
    1 StVenantTorsion_Fe_Tri6
        NodalDof 1
        Stage 1
        Subsystem 1
        CellFieldOutput 4
            1 s_zx
            2 s_zy
            3 s_mag
            4 u_z
        TwistPerLength 0.003
        DofGroup 1

*MATERIALS 2
    1 LinearIsotropicElasticity Torsion 100
    2 LinearIsotropicElasticity Torsion 10

*DOMAIN_ASSIGNMENTS 3
    "LowerBlock" Numerics 1 MaterialSet 1
    "InnerAngle" Numerics 1 MaterialSet 2
    "HollowInsert" Numerics 1 MaterialSet 1

*OUTPUT_FORMAT Gmsh
    FILENAME  Torsion
    OUTPUT_STYLE ascii
    NODE_DATA 1
        SCALAR WarpingFunction 1
    ELEMENT_DATA 0
    ELEMENT_NODE_DATA 5
        SCALAR Stress_zx 1
        SCALAR Stress_zy 2
        SCALAR Stress_magnitude 3
        VECTOR StressVector 1 2 0
        SCALAR Z_Displacement 4

*CSV_OUTPUT 0

*LOADSTEPS 1
    1 PREPROCESSING 0
       START_TIME 0.0
       END_TIME 1.0
       INITIAL_TIME_INCREMENT 1.0
       MAX_SUBSTEPS 100
       BOUNDARY_CONDITIONS	1
           "LateralSurface" 1 TorsionBoundaryTerm 1 None
       FIELD_CONDITIONS 0
       SOLUTION_METHODS
           Stage 1 LinearStatic
               LinearSolver MKL_Pardiso nThreads 12 Unsymmetric
       WRITE_INTERVAL 0
       POSTPROCESSING 0
*END
\end{lstlisting}
In particular, the implementation defines only one type of boundary condition, \texttt{TorsionBoundaryTerm}. This automatically distinguishes between material interfaces and free surface boundaries depending on whether the lower-dimensional boundary element (3-node curve in this case) has one or two adjacent domain (6-node) elements. The distribution of shear stress components $\sigma_{zx}$ and $\sigma_{zy}$ are plotted in Figures \ref{fig:stressContinuityZX} and \ref{fig:stressContinuityZY}. One can observe that $\sigma_{zx}$ varies continuously across vertical material interfaces while $\sigma_{zy}$ is continuous across horizontal interfaces, in accordance with \eqref{eq:stVenantInternalBoundary}. Meanwhile, the effect of relative rigidity between subdomains on the distribution and flow of shear stresses is shown on Figure \ref{fig:ratioEffect}.
We can see that as the relative stiffness of one subdomain becomes negligible compared to its counterpart across a material interface, the behavior of said internal boundary shifts towards that of a free surface which in turn affects the flow of shear.

\subsection{Monolithic FE-FV formulation for Biot poroelasticity}
\begin{figure}
	\centering
	\begin{subfigure}[t]{0.33\textwidth}
		\includegraphics[width=\textwidth]{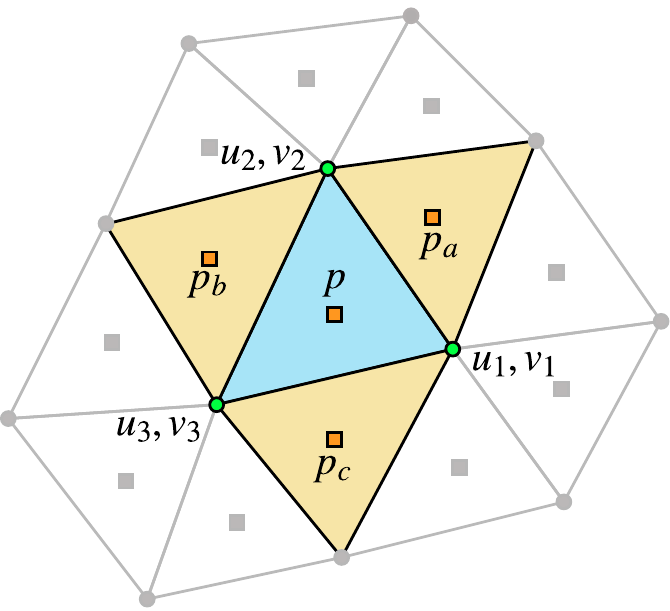}
		\caption{}
	\end{subfigure}
	\begin{subfigure}[t]{0.65\textwidth}
		\includegraphics[width=\textwidth]{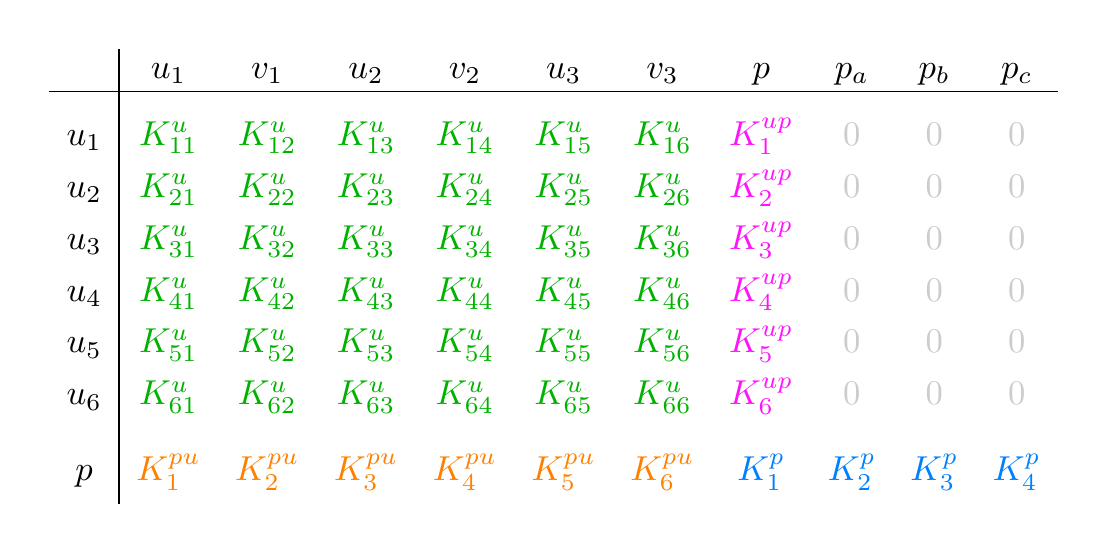}
		\caption{}
	\end{subfigure}
	\caption{Local stencil for combined FE-FV discretization of the poroelasticity problem and corresponding local coefficient matrix for a fully coupled solution.}
	\label{fig:stencilPoroelasticity}
\end{figure}

The linear poroelasticity theory of Biot\cite{Biot1941} is a popular choice for describing the coupled response between mechanical deformation and fluid flow in a porous medium. It is based on the following assumptions:
\begin{enumerate}[a)]
	\setlength{\itemsep}{0pt}
	\setlength{\parskip}{0pt}
	\setlength{\parsep}{0pt}
	\item the medium is saturated with a single-phase fluid,
	\item Darcy's Law is valid,
	\item mechanical deformation is quasi-static and results in infinitesimal strains, and
	\item the porous medium skeleton exhibits linear elastic behavior.
\end{enumerate}
The unknown fields are the displacement $\mathbf{u}$ of the porous medium skeleton, and the fluid pressure $p$. Meanwhile the governing equations consist of two PDEs: a linear momentum conservation equation for the porous medium and mass conservation equation for the fluid. For some given domain $\Omega$ having boundary $\partial\Omega$, the complete initial-boundary value problem can be expressed as
\begin{linenomath}
\begin{numcases}{}
	\nabla \cdot \left( \mathbf{\sigma}^\prime - \alpha p \mathbf{I} \right) + \rho_\text{sat} \mathbf{g} = \mathbf{0} & in $\Omega$ \\
	\mathbf{u} = \bar{\mathbf{u}} & on $\partial\Omega_\mathbf{u}^D$ \\
	\left(\mathbf{\sigma}^\prime - \alpha p \mathbf{I} \right) \cdot \mathbf{n} = \bar{\mathbf{t}} & on $\partial\Omega_\mathbf{u}^N$ \\
	\alpha \frac{\partial\epsilon_v}{\partial t} + \frac{1}{M} \frac{\partial p}{\partial t} + \nabla \cdot \left( -\frac{\mathbf{k}}{\mu} \cdot \nabla p \right) = Q & in $\Omega$ \\
	p = \bar{p} & on $\partial\Omega_p^D$ \\
	-\frac{\mathbf{k}}{\mu} \cdot \nabla p = \bar{\mathbf{q}} & on $\partial\Omega_p^N$ \\
	p \left( \mathbf{x}, 0 \right) = \bar{p} \left( \mathbf{x} \right) & in $\Omega$
\end{numcases}
\end{linenomath}
wherein the quantity $\mathbf{\sigma}^\prime = \mathbb{C} : \mathbf{\epsilon} \left( \mathbf{u} \right)$ is termed the \emph{effective stress}, and $\mathbf{\sigma} = \mathbf{\sigma}^\prime - \alpha p \mathbf{I}$ the \emph{total/poroelastic stress}. The adopted sign convention assumes that positive values for normal stresses denote tension, whereas fluid pressure is taken positive in compression.  For pure solid mechanics problems, the most common approach is to express the equilibrium equation in a weak/variational form and to employ a method such as finite elements, as the resulting secondary variables have a well defined physical meaning. On the other hand, it is often preferred to solve the primal formulation of the fluid mass balance equation with a method such as finite volumes that enforces local mass conservation across cell boundaries. In contrast, continuous Galerkin methods enforce node-wise conservation of element nodal fluxes\cite{Hughes2000}. In the following two examples, we combine a nodal finite element scheme for stress equilibrium with a cell-centered finite volume method for fluid flow. The former utilizes piecewise contiuous $P_1$ shape functions constructed over a simplex mesh, while the latter makes use of the two-point flux approximation scheme (TPFA) to construct the fluxes across simplex element boundaries. The resulting FE-FV stencil is depicted in Figure \ref{fig:stencilPoroelasticity} together with the local coefficient matrix profile corresponding to a fully coupled approach.
Notably, the latter is not square and furthermore has around 1/4 of its total entries equal to zero. One can readily see that for these type of problems, passing local contributions to the global matrix assembler in triplet (COO) format is preferable to a standard rectangular matrix since it allows for exclusion of local components that are known to be identically zero. Otherwise, the resulting nonzero profile of the global coefficient matrix becomes denser than what is necessary, which would in turn adversely affect the performance of linear solvers.

\subsubsection{Mandel's problem}
\begin{figure}
	\centering
	\begin{subfigure}[b]{0.4\textwidth}
		\includegraphics[width=\textwidth]{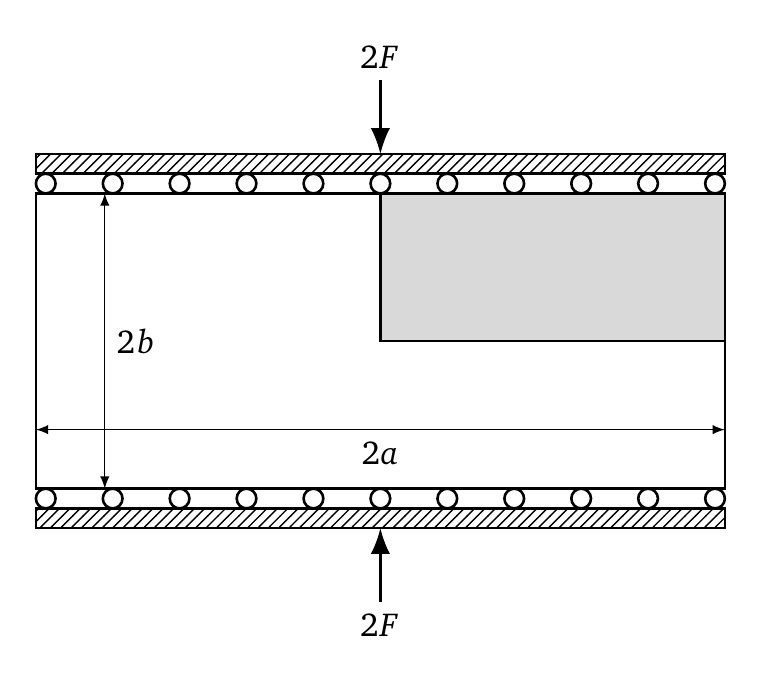} \\[-25pt]
		\caption{}
	\end{subfigure} \hspace{10pt}
	\begin{subfigure}[b]{0.45\textwidth}
		\includegraphics[width=\textwidth]{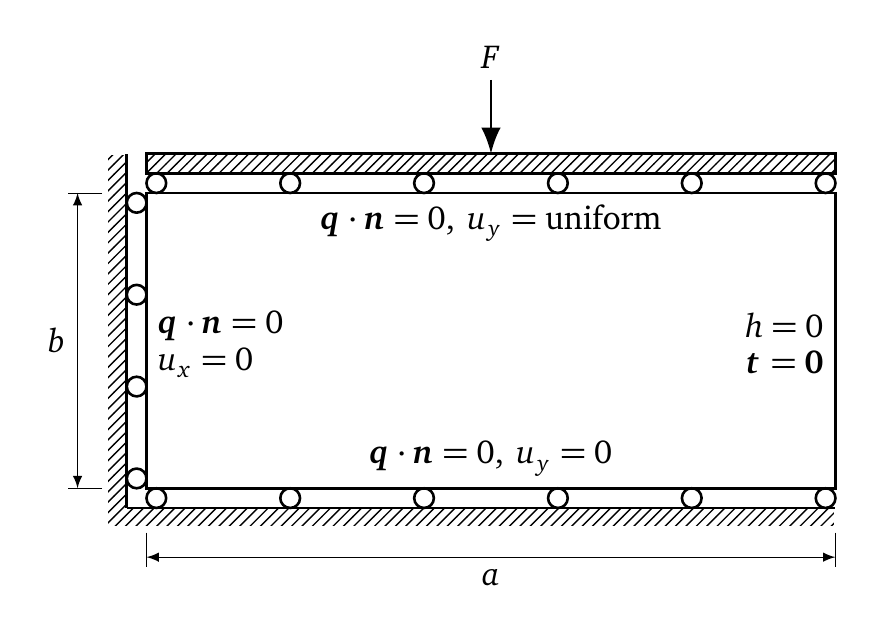}
		\caption{}
	\end{subfigure}
	\caption{Setup for Mandel's problem, showing (a) full geometry, and (b) computational domain.}
	\label{fig:mandel}
\end{figure}

Mandel's problem \cite{Mandel1953} is a popular benchmark problem for testing the performance of both discretization schemes and solution algorithms owing to the fact that it is a non-trivial time-dependent problem in 2D with a known analytical solution. The set-up is shown in Figure \ref{fig:mandel} and involves the squeezing of a saturated poroelastic slab between two rigid, frictionless and impervious plates.
The slab and plates are assumed to extend infinitely along the out-of-plane ($z$) direction so that deformation follows plane strain conditions, i.e., $u_z = 0$, and $u_x$ and $u_y$ are independent of $z$. Due to the problem symmetry, it suffices to consider only the quarter domain $\left( 0,a \right) \times \left( 0,b \right)$, for which the relevant boundary conditions are as follows:
\begin{linenomath}
	\begin{subequations}
		\begin{align}
		u_x &= 0 \text{ on } x = 0 \\
		u_y &= 0 \text{ on } y = 0 \\
		u_y &= \bar{u} \left( t \right) \text{ on } y = b \label{eq:mandelRigidConstraint} \\
		\int_0^a \left( \sigma^\prime_{yy}  - p \right) \dee x &= -F \text{ on } y = b  \label{eq:mandelAppForce} \\
		\sigma_{xy} = \sigma_{xx} = p &=0 \text{ on } x = a \\
		\left( -\frac{\mathbf{k}}{\mu} \cdot \nabla p \right) \cdot \mathbf{n} &= 0 \text{ on } x = 0, y = 0 \text{ and } y = b.
		\end{align}
	\end{subequations}
\end{linenomath}
In addition, initial conditions are set to zero for both $\mathbf{u}$ and $p$. Rigidity of the plates along with problem symmetry implies a uniform vertical displacement $\bar{u}$ at $y = b$ that is a priori unknown but dependent only on time. To achieve this condition, one must be able to constrain all degrees of freedom pertaining to $u_y$ on the top boundary to have the same value. {\broomstyx} allows the end user to specify master/slave types of constraints in the input file, by selecting one master DOF (e.g., the one residing on the node located at $\left( 0,b \right)$) and declaring all other $u_y$ DOFs on $y = b$ as slaves. The constraint is enforced in a strong sense during the simulation: unknowns associated with the slave DOFs are eliminated from the system of equations prior to determining the global sparse matrix structure. This also lumps the right hand side terms (i.e.\ resultant forces) onto the master DOF, which corresponds to the integral-type condition given in \eqref{eq:mandelAppForce}.

\begin{figure}
	\centering
	\includegraphics[width=0.5\textwidth]{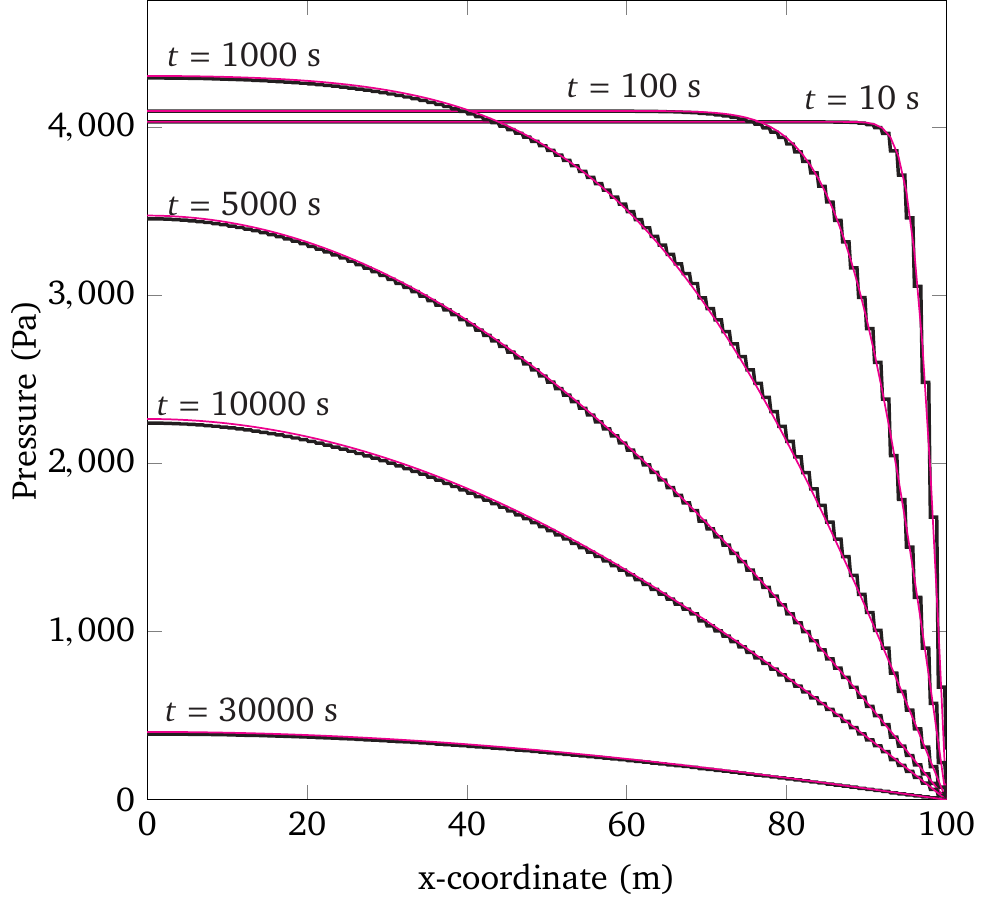}
	\caption{Analytical vs. numerical results for the Mandel problem.}
	\label{fig:mandel_results}
\end{figure}

Analytical expressions for the displacements and stresses have been derived by Abousleiman et al.\cite{Abousleiman1996}, supplementing the original solution of Mandel\cite{Mandel1953} for the pressure. Due to the instantaneous application of external loading, the solutions are time-discontinuous at $t = 0$. Material properties for the current simulation are taken from the paper of Mikeli\'c et al.\cite{Mikelic2014} and are listed in Table \ref{tab:MandelParameters}.
\begin{table}
	\centering
	\caption{Dimensions and material parameters for Mandel's problem.}
	\label{tab:MandelParameters}
	\begin{tabular}{lll}
		\toprule Symbol & Description & Value \\
		\midrule $a$ & Horizontal dimension & 100 m \\
		$b$ & Vertical dimension & 10 m \\
		$E$ & Drained Young's modulus & $5.94 \times 10^9$ Pa \\
		$\nu$ & Drained Poisson ratio & 0.2 \\
		$k$ & Intrinsic permeability & $1.0 \times 10^{-13}$ m$^2$ \\
		$\mu$ & Dynamic fluid viscosity & 1.0 cP \\
		$\alpha$ & Biot-Willis coefficient & 1.0 \\
		$C_f$ & Fluid compressibility & $3.03 \times 10^{-10}$ Pa$^{-1}$ \\
		\bottomrule
	\end{tabular}		
\end{table}
A comparison of analytical and numerical solutions is shown in Figure \ref{fig:mandel_results}.
We can observe that the combined FE-FV scheme yields numerical results of the pressure that are in very good agreement with the analytical solution.

\subsubsection{Terzaghi 1-D consolidation}
\begin{figure}
	\centering
	\begin{subfigure}[b]{0.4\textwidth}
		\includegraphics[width=\textwidth]{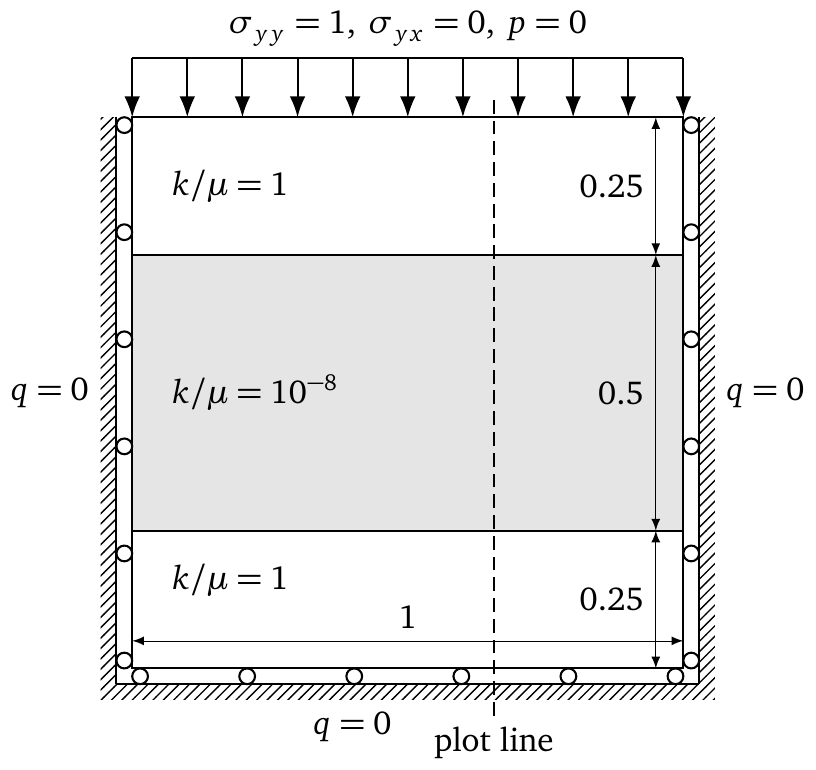}
		\caption{}
		\label{fig:layered_medium_geometry}
	\end{subfigure} \hspace{1em}
	\begin{subfigure}[b]{0.3\textwidth}
		\includegraphics[width=\textwidth]{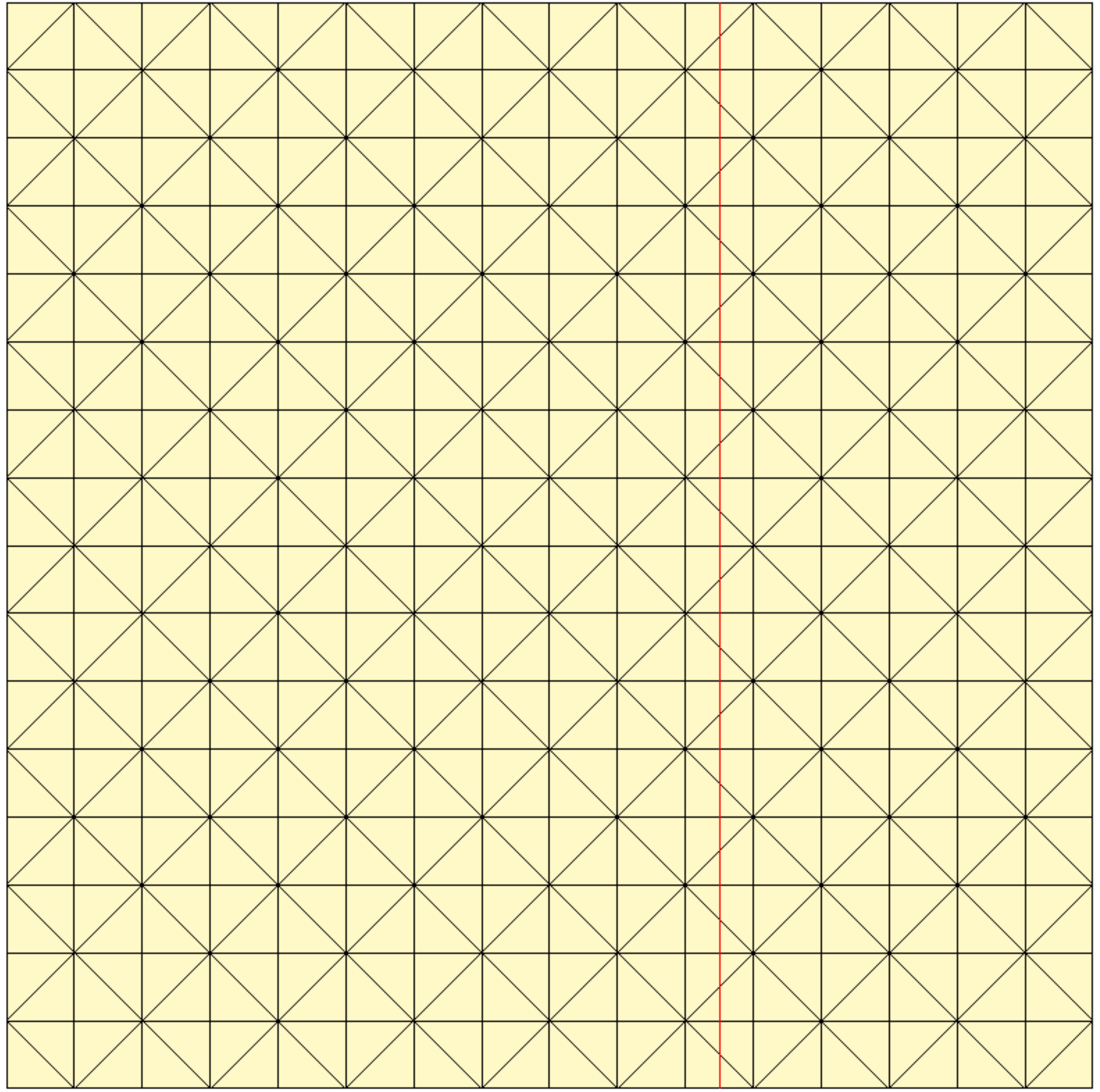} \vspace{1em}
		\caption{}
		\label{fig:layered_medium_mesh}
	\end{subfigure}
	\caption{Layered porous medium undergoing consolidation. Geometry, material properties and boundary conditions shown in (a). The domain is discretized using a structured mesh of alternating triangles as shown in (b);  with the plot line located at $x = 0.65625$.}
	\label{fig:layered_medium}
\end{figure}

\begin{figure}
	\centering
	\includegraphics[width=0.45\textwidth]{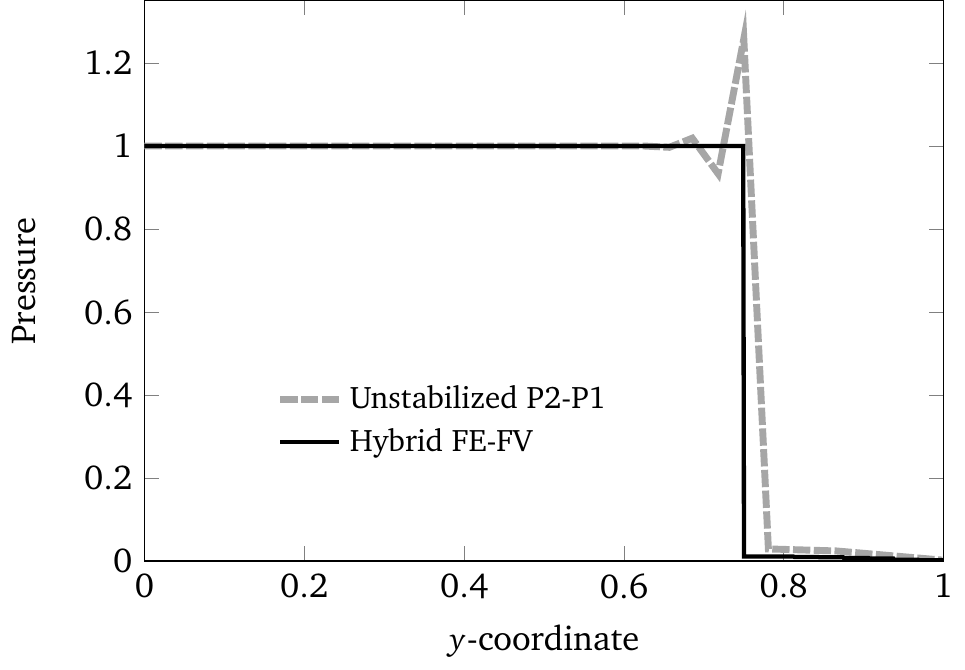}
	\caption{Vertical variation of pressure at $t = 1$ along the plot line (See Figure \ref{fig:layered_medium}) in the layered porous medium, obtained via the hybrid FE-FV scheme. Results reported by Rodrigo et al. \cite{Rodrigo2016} for unstabilized $P_2$-$P_1$ elements are shown for comparison.}
	\label{fig:layered_porous_medium_results}
\end{figure}

The consolidation of a layered porous medium under a uniform external distributed load was initially proposed in Haga et al. \cite{Haga2012}, and later utilized by Rodrigo et al. \cite{Rodrigo2016} to demonstrate the occurrence of numerical oscillations in finite element solutions for the pressure at early times even when inf-sup stable elements are used. The problem domain is a square having each side of unit measure and consisting of a middle region of low permeability sandwiched between two highly permeable layers as illustrated in Figure \ref{fig:layered_medium_geometry}.
The medium is assumed to be saturated an incompressible fluid ($C_f = 0$). For all layers, the Lam\'e parameters as well the Biot-Willis coefficient are set to unity, i.e.\ $\lambda = \mu = \alpha = 1$. On the other hand, the mobility coefficient $k/\mu$ is set to 1 in the outer layers and $10^{-8}$ in the middle layer. Zero initial conditions are assumed for both $\mathbf{u}$ and $p$, and the time step is chosen as $\Delta t = 1$. For the current example, the domain is discretized using a structured triangular mesh as shown in Figure \ref{fig:layered_medium_mesh}, with numerical results for the pressure at $t = 1$ along a vertical line situated 0.65625 units from the left side of the domain. For comparison, these are plotted in Figure \ref{fig:layered_porous_medium_results} together with the numerical solution obtained using unstabilized Taylor-Hood ($P_2$-$P_1$) elements.
As the pressure is piecewise constant across elements in the FV scheme used to model the flow equation, the initial jump in pressure at the interface between the upper and middle layers can be properly reproduced without the occurrence of overshoots, in contrast with formulations that model pressure as a continuous variable in space. Nonetheless slight oscillations are observed at later times. In particular for the current set of material properties, the oscillations achieve maximum magnitude at around $t = 3400$ (in a classical checkerboard pattern) and gradually disappear thereafter.

\subsection{Brittle fracture of a heterogeneous specimen}
Phase-field models for fracture are a relatively recent development in the direction of diffuse approaches to fracture modeling that have garnered a lot of interest in the last decade. In the phase-field method, it is assumed that the total potential energy associated with a fracturing solid medium is composed of bulk and surface terms along with external work terms, and can be regularized using a scalar damage/phase-field variable $\phi$ to yield the functional\cite{Bourdin2008}
\begin{linenomath}
\begin{equation}
	\Pi \left( \mathbf{u}, \phi \right) = \underbrace{\int_\Omega \psi \left( \mathbf{\epsilon}, \phi \right) \dee\Omega}_\text{Bulk energy}  - \int_\Omega \mathbf{b} \cdot \mathbf{u} \,\dee\Omega - \int_{\partial\Omega^N} \mathbf{t} \cdot \mathbf{u} \,\dee\Gamma + \underbrace{\mathcal{G}_c \int_\Omega \left( \frac{1}{2\ell} \phi^2 + \frac{\ell}{2} \nabla\phi \cdot \nabla\phi \right) \dee\Omega}_\text{Surface energy} ,
	\label{eq:phaseFieldFunctional}
\end{equation}
\end{linenomath}
where $\mathbf{u}$ and $\mathbf{\epsilon} = \mathbf{\epsilon} \left( \mathbf{u} \right)$ denote respectively the displacement field and infinitesimal strain tensor, $\mathbf{b}$ is the body force, and $\mathbf{t}$ is the external traction acting on the Neumann boundary $\partial\Omega^N$. On the other hand, $\mathcal{G}_c$ is the critical energy release rate from Griffith's theory of brittle fracture and $\ell$ is a length scale controlling the amount of regularization/diffusion in the resulting crack representation. The functional in \eqref{eq:phaseFieldFunctional} leads to the classical 2nd-order phase-field model that is utilized in the current example. The regularized bulk energy density $\psi \left( \mathbf{\epsilon}, \phi \right)$ can be either isotropic or anisotropic, i.e.
\begin{linenomath}
\begin{subequations}
\begin{align}
	&\psi \left( \mathbf{\epsilon}, \phi \right) = g \left( \phi \right) \psi_0 \left( \mathbf{\epsilon} \right) \hspace{6em} \text{(isotropic model)} \label{eq:bulkDensityIsotropic} \\[0.5em]
	&\psi \left( \mathbf{\epsilon}, \phi \right) = g \left( \phi \right) \psi_0^+ \left( \mathbf{\epsilon} \right) + \psi_0^- \left( \mathbf{\epsilon} \right)  \hspace{2em} \text{(anisotropic model)}
	\label{eq:bulkDensityAnisotropic}
\end{align}
\end{subequations}
\end{linenomath}
where $\psi_0 \left( \mathbf{\epsilon} \right)$ is the elastic strain energy density, and $g \left( \phi \right)$ is an energy degradation function. Imposing stationarity of \eqref{eq:phaseFieldFunctional} with respect to both $\mathbf{u}$ and $\phi$ yields the coupled system of PDEs describing the brittle fracture evolution problem, expressed in weak form as
\begin{linenomath}
\begin{numcases}{}
	\int_\Omega \frac{\partial \psi}{\partial\mathbf{\epsilon}} : \delta\mathbf{\epsilon} \,\dee\Omega = \int_\Omega \mathbf{u} \cdot \delta\mathbf{\epsilon} \,\dee\Omega + \int_{\partial\Omega^N} \mathbf{t} \cdot \delta\mathbf{u} \,\dee\Gamma \label{eq:weakForm_mechanics} \\[0.5em]
	\int_\Omega \frac{\partial\psi}{\partial\phi} \delta\phi \,\dee\Omega + \int_\Omega \mathcal{G}_c \left( \ell \nabla\phi \cdot \nabla\delta\phi + \frac{1}{\ell} \phi \, \delta\phi \right) \dee\Omega = 0
	\label{eq:weakForm_phaseField}
\end{numcases}
\end{linenomath}
Owing to the form of $\psi \left( \mathbf{\epsilon}, \phi \right)$, the above system represents a nonlinear set of equations that must be solved via an approach that can handle the non-convexity of \eqref{eq:phaseFieldFunctional}. In {\broomstyx}, \eqref{eq:weakForm_mechanics}--\eqref{eq:weakForm_phaseField} is implemented via the class \class{PhaseFieldFracture\_Fe\_Tri3} which assumes 2D plane strain conditions. Furthermore, $\psi \left( \mathbf{\epsilon}, \phi \right)$ is implemented as a \class{Material} class through a composition strategy, in that it contains among its members additional material objects corresponding to elasticity models and degradation functions. This maximizes code reuse, since several different models for $\psi \left( \mathbf{\epsilon}, \phi \right)$ have been proposed in the literature. Here we make use of the anisotropic model of Amor et al.\cite{Amor2009} that enforces the unilateral contact condition through a volumetric/deviatoric splitting of the energy. Let $\mathbf{\epsilon} = \left\{ \epsilon_{xx}, \epsilon_{yy}, 0, \gamma_{xy} \right\}^\trp$ be the symmetric strain tensor in Voigt form.
\footnote{It is assumed that constitutive models return the elasticity tensor as a $3 \times 3$ matrix when in plane stress mode, $4 \times 4$ in plane strain and axisymmetric analysis, and $6 \times 6$ for the full 3D case; the use of a rank-4 matrix in plane strain is due to the fact that the stress component $\sigma_{zz}$ is needed when dealing with plasticity. However since $\epsilon_{zz} = 0$ in plane strain, the symmetrized gradient matrix $\mathbf{B}$ that is used to obtain strains from displacements via the operation $\mathbf{\epsilon} = \mathbf{B} \mathbf{u}$ has hard-coded zeros corresponding to the row for $\epsilon_{zz}$, i.e.
\begin{equation}
	\mathbf{B} = \left[ \begin{array}{ccccc} N_{1,x} & 0 & N_{2,x} & 0 & \ldots \\[0.4em] 0 & N_{1,y} & 0 & N_{2,y} & \ldots \\[0.4em] 0 & 0 & 0 & 0 & \ldots \\[0.2em] N_{1,y} & N_{1,x} & N_{2,y} & N_{2,x} & \ldots \end{array} \right]
\end{equation}
\label{ftnote:bmat}
}
Its volumetric and deviatoric components can then be calculated as
\begin{linenomath}
\begin{equation}
	\mathbf{\epsilon}_\text{vol} = \mathbb{P} \mathbf{\epsilon} \qquad \mathbf{\epsilon}_\text{dev} = \left( \mathbb{I} - \mathbb{P} \right) \mathbf{\epsilon}
\end{equation}
\end{linenomath}
wherein
\begin{linenomath}
\begin{equation}
	\mathbb{I} = \left[ \begin{array}{m{1em}m{1em}m{1em}c} 1 & 0 & 0 & 0 \\ 0 & 1 & 0 & 0 \\ 0 & 0 & 1 & 0 \\ 0 & 0 & 0 & 1 \end{array} \right] \text{ and }
	\mathbb{P} = \frac{1}{2}\left[ \begin{array}{m{1em}m{1em}m{1em}c} 1 & 1 & 0 & 0 \\ 1 & 1 & 0 & 0 \\ 0 & 0 & 0 & 0 \\ 0 & 0 & 0 & 0 \end{array} \right] ,
\end{equation}
\end{linenomath}
with the given form of $\mathbb{P}$ being specific for 2D analysis.
\footnote{The decomposition of $\mathbf{\epsilon}$ into volumetric and deviatoric components is normally written as $\mathbf{\epsilon} = \dfrac{1}{n} \epsilon_v \mathbf{I} + \mathbf{\epsilon}_\text{dev}$, in which $\epsilon_v = \mathrm{tr} \left( \mathbf{\epsilon} \right)$ and $n = 3$, so that $\mathrm{tr} \left( \mathbf{\epsilon}_\text{dev} \right) = 0$. However in plane strain, $\epsilon_{zz}$ is identically zero hence $\epsilon_v = \epsilon_{xx} + \epsilon_{yy}$. For the current problem, we perform the decomposition assuming $n = 2$, such that in the 3rd dimension $\epsilon_{zz,\text{vol}} = \epsilon_{zz,\text{dev}} = 0.$ This of course means that $\mathbf{\epsilon}_\text{vol}$ is no longer isotropic in a full 3D context.}
Then for some strain energy function $\psi_0 \left( \phi \right)$, we obtain
\begin{linenomath}
\begin{equation}
	\psi_{0,\text{vol}} \left( \mathbf{\epsilon} \right) = \psi_0 \left(  \mathbf{\epsilon}_\text{vol} \right) \qquad \psi_{0,\text{dev}} \left( \mathbf{\epsilon} \right) = \psi_0 \left(  \mathbf{\epsilon}_\text{dev} \right)
\end{equation}
\end{linenomath}
Likewise, the stress decomposition is given by
\begin{linenomath}
\begin{equation}
	\mathbf{\sigma}_\text{vol} = \frac{\dee\psi_0 \left( \mathbf{\epsilon}_\text{vol} \right)}{\dee\mathbf{\epsilon}} \qquad \mathbf{\sigma}_\text{dev} = \frac{\dee\psi_0 \left( \mathbf{\epsilon}_\text{dev} \right)}{\dee\mathbf{\epsilon}}
\end{equation}
\end{linenomath}
On the other hand, the derivatives of $\mathbf{\sigma}_\text{vol}$ and $\mathbf{\sigma}_\text{dev}$ with respect to the strain are
\begin{linenomath}
\begin{equation}
	\mathbb{C}_\text{vol} = \frac{\dee\mathbf{\sigma}_\text{vol}}{\dee\mathbf{\epsilon}} =  \frac{\dee^2\psi_0 \left( \mathbf{\epsilon}_\text{vol} \right)}{\dee\mathbf{\epsilon} \,\dee\mathbf{\epsilon}} \mathbb{P} \qquad \mathbb{C}_\text{dev} = \frac{\dee\mathbf{\sigma}_\text{dev}}{\dee\mathbf{\epsilon}} = \frac{\dee^2\psi_0 \left( \mathbf{\epsilon}_\text{dev} \right)}{\dee\mathbf{\epsilon} \,\dee\mathbf{\epsilon}} \left( \mathbb{I} - \mathbb{P} \right),
\end{equation}
\end{linenomath}
both of which are asymmetric. Nevertheless, the products $\mathbf{B}^\trp \mathbb{C}_\text{vol} \mathbf{B}$ and $\mathbf{B}^\trp \mathbb{C}_\text{dev} \mathbf{B}$ yield symmetric matrices due to the particular form of $\mathbf{B}$ in plane strain (see footnote \ref{ftnote:bmat}). In Amor et al.\cite{Amor2009}, the damage-degraded bulk energy density is given as
\begin{linenomath}
\begin{equation}
	\psi \left( \mathbf{\epsilon}, \phi \right) = g \left( \phi \right) \left[ \psi_0 \left( \mathbf{\epsilon}_\text{vol}^+ \right) + \psi_0 \left( \mathbf{\epsilon}_\text{dev} \right) \right] + \psi_0 \left( \mathbf{\epsilon}_\text{vol}^- \right),
\end{equation}
\end{linenomath}
wherein $\mathbf{\epsilon}_\text{vol}^+ = H \left( \trace\mathbf{\epsilon} \right) \mathbf{\epsilon}_\text{vol}$ and $\mathbf{\epsilon}_\text{vol}^- = H \left( -\trace\mathbf{\epsilon} \right) \mathbf{\epsilon}_\text{vol}$ with $H \left( \bullet \right)$ denoting the Heaviside step function. This is implemented in the material class \class{AmorDamageModel}, which as previously mentioned requires also the specification of further classes to calculate quantities related to the basic elasticity model as well as the energy degradation function, for instance as illustrated in Table \ref{tab:overriddenMaterialMethods1}.
\begin{table}
	\centering
	\caption{Quantities returned by overridden methods for implementations of derived \class{Material} classes corresponding to a isotropic, linear elastic solid, and the standard quadratic degradation function used for fracture phase-field models.}
	\label{tab:overriddenMaterialMethods1}
	\begin{tabular}{lccc}
		\toprule
		\multicolumn{1}{c}{Class method} & Return type & \multicolumn{2}{c}{Returned quantity} \\ \cmidrule{3-4} 
		&& \class{LinearIsotropicElasticity} & \class{QuadraticDegradation} \\
		\midrule
		\texttt{givePotentialFrom(...)} & \texttt{double} & $\psi_0 \left( \mathbf{\epsilon} \right) = \dfrac{1}{2} \mathbf{\epsilon} : \mathbb{C} : \mathbf{\epsilon}$ & $g \left( \phi \right) = \left( 1 - \phi \right)^2 $ \\
		\texttt{giveForceFrom(...)} & \class{RealVector} & $\dfrac{\dee \psi_0}{\dee \mathbf{\epsilon}} = \mathbb{C} : \mathbf{\epsilon}$ & $g^\prime \left( \phi \right) = -2 \left( 1 - \phi \right)$ \\[0.7em]
		\texttt{giveModulusFrom(...)} & \class{RealMatrix} & $\dfrac{\dee^2 \psi_0}{\dee \mathbf{\epsilon} \,\dee \mathbf{\epsilon}} = \mathbb{C}$ & $g^{\prime\prime} \left( \phi \right) = 2$ \\[0.5em] \bottomrule
	\end{tabular}
\end{table}
The parent material model makes the appropriate calls to its child objects to compute the necessary intermediate quantities; these are then combined together to give the requested potential, force or modulus quantity as detailed in Table \ref{tab:AmorModel}.
\begin{table}
	\centering
	\caption{Quantities returned by polymorphic methods for derived class \class{AmorDamageModel}, which are built up from intermediate results obtained from calling methods of child objects (cf. Table \ref{tab:overriddenMaterialMethods1}) with the constitutive state vector set to $\mathbf{\epsilon}_\text{vol}^+$, $\mathbf{\epsilon}_\text{vol}^-$, $\mathbf{\epsilon}_\text{dev}$ or $\phi$ as appropriate.}
	\label{tab:AmorModel}
	\begin{tabular}{lc}
		\toprule
		\multicolumn{1}{c}{Class method} & Returned quantity \\ \midrule
		\texttt{givePotentialFrom(...)} & $\psi \left( \mathbf{\epsilon}, \phi \right) = g \left( \phi \right) \left[ \psi_0 \left( \mathbf{\epsilon}_\text{vol}^+ \right) + \psi_0 \left( \mathbf{\epsilon}_\text{dev} \right) \right] + \psi_0 \left( \mathbf{\epsilon}_\text{vol}^- \right)$ \\[0.5em]
		\texttt{giveForceFrom(...,"Mechanics")} & $\dfrac{\dee \psi}{\dee \mathbf{\epsilon}} = g \left( \phi \right) \left[ \dfrac{\dee \psi_0 \left( \mathbf{\epsilon}_\text{vol}^+ \right)}{\dee \mathbf{\epsilon}} + \dfrac{\dee \psi_0 \left( \mathbf{\epsilon}_\text{dev} \right)}{\dee \mathbf{\epsilon}} \right] + \dfrac{\dee \psi_0 \left( \mathbf{\epsilon}_\text{vol}^- \right)}{\dee \mathbf{\epsilon}}$ \\[1.5em]
		\texttt{giveModulusFrom(...,"Mechanics")} & $\dfrac{\dee^2 \psi}{\dee \mathbf{\epsilon} \,\dee \mathbf{\epsilon}} = g \left( \phi \right) \left[ \dfrac{\dee^2 \psi_0 \left( \mathbf{\epsilon}_\text{vol}^+ \right)}{\dee \mathbf{\epsilon} \,\dee \mathbf{\epsilon}} \mathbb{P} + \dfrac{\dee^2 \psi_0 \left( \mathbf{\epsilon}_\text{dev} \right)}{\dee \mathbf{\epsilon} \,\dee \mathbf{\epsilon}} \left( \mathbb{I} - \mathbb{P} \right) \right] + \dfrac{\dee^2 \psi_0 \left( \mathbf{\epsilon}_\text{vol}^- \right)}{\dee \mathbf{\epsilon} \,\dee \mathbf{\epsilon}} \mathbb{P}$ \\[1em]
		\texttt{giveForceFrom(...,"PhaseField")} & $\dfrac{\dee \psi}{\dee\phi} = g^\prime \left( \phi \right) \left[ \psi_0 \left( \mathbf{\epsilon}_\text{vol}^+ \right) + \psi_0 \left( \mathbf{\epsilon}_\text{dev} \right) \right]$ \\[1em]
		\texttt{giveModulusFrom(...,"PhaseField")} & $\dfrac{\dee \psi}{\dee\phi} = g^{\prime\prime} \left( \phi \right) \left[ \psi_0 \left( \mathbf{\epsilon}_\text{vol}^+ \right) + \psi_0 \left( \mathbf{\epsilon}_\text{dev} \right) \right]$ \\[1em]
		\bottomrule
	\end{tabular}
\end{table}

\begin{figure}
	\centering
	\includegraphics[width=0.25\textwidth]{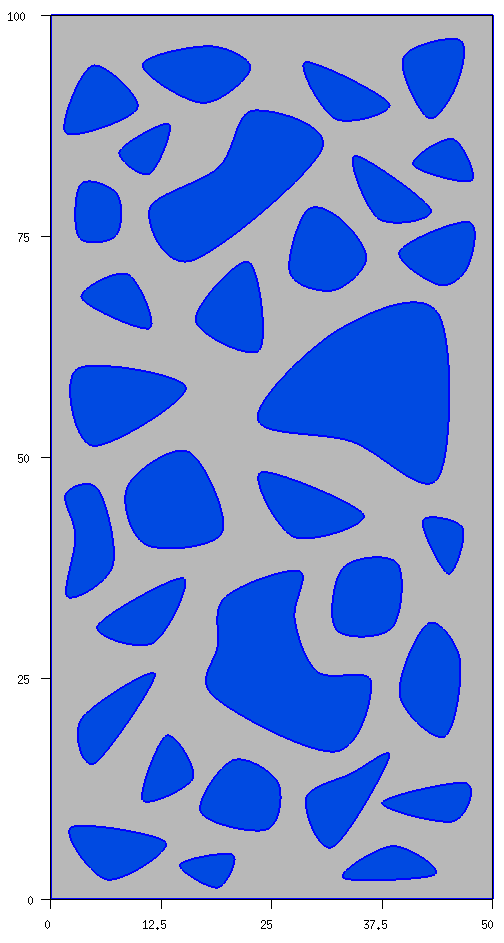}
	\caption{Concrete specimen in unconfined compression test, consisting of aggregate material (blue) embedded in a cement paste matrix (gray).}
	\label{fig:concreteSpecimen}
\end{figure}

In the current example, we make use of the classes described above to simulate shear failure in rectangular concrete specimen in an unconfined compression test. The height and width of said specimen are 100 mm and 50 mm respectively, and it is assumed to be made up of a cement paste matrix that contains inclusions representing aggregate material as shown in Figure \ref{fig:concreteSpecimen}.
To demonstrate the framework's capability to handle configurations of the type illustrated in Figure \ref{fig:domainConfig3}, we assume that only the cement paste can experience fracture, while the inclusions behave in a purely elastic manner. Table \ref{tab:materialProperties_UCTest} gives a summary of the material properties for the different components of the specimen.
\begin{table}
	\centering
	\caption{Material properties for unconfined compression test.}
	\label{tab:materialProperties_UCTest}
	\begin{tabular}{lcccc}
		\toprule
		& Young's modulus & Poisson ratio & Tensile strength & Critical energy release rate \\
		& $E$ (GPa) & $\nu$ & $f_t$ (MPa) & $\mathcal{G}_c$ ( J/m$^2$) \\
		\midrule
		Cement paste\cite{Brown1973,Haecker2005,Li2015} & 30 & 0.25 & 3.1 & 2.8 \\
		Inclusions\cite{Schultz1995} & 77 & 0.2 & --- & --- \\
		\bottomrule
	\end{tabular}
\end{table}
In addition, no debonding is allowed at the interfaces so that fractures must occur completely within the cement matrix.  Perfect bonding is assumed between the aggregate inclusions and the surrounding cement matrix, hence the displacement field is assumed to be continuous across material interfaces. Hence for the linear momentum equation, the relevant boundaries consist of the external sides of the specimen. In particular, its left sides are assumed to be traction free, while vertical compressive loading is applied through its top and bottom surfaces which are assumed to be smooth. This translates to the following essential boundary conditions: $u_y \left( x, 0 \right) = 0$, $u_y \left( x, 100 \right) = U$ and $u_x \left( 0, 0 \right) = 0$ with the last condition needed to ensure uniqueness of solutions. The final prescribed displacement at the top surface of the specimen is set to $U = -0.0125$ mm, which is applied incrementally, initially with $\Delta u_y = 5 \times 10^{-4}$ mm and refined to $\Delta u_y = 1 \times 10^{-4}$ near the instance of failure. On the other hand for the phase-field equation, the relevant boundaries consist of the external surfaces of the specimen together with the interfaces between cement matrix and inclusions. The condition $\nabla\phi \cdot \mathbf{n} = 0$ is imposed on the external surfaces of the specimen, while at material interfaces, it is possible to impose either $\nabla\phi \cdot \mathbf{n} = 0$ or $\phi = 0$. We use the standard quadratic degradation function
\begin{linenomath}
\begin{equation}
	g \left( \phi \right) = \left( 1 - \phi \right)^2,
\end{equation}
\end{linenomath}
and following Nguyen et al.\cite{Nguyen2016} the value of the phase-field regularization parameter is calculated as
\begin{linenomath}
	\begin{equation}
	\ell = \frac{27E\mathcal{G}_c}{256 f_t^2}.
	\end{equation}
\end{linenomath}
Plugging in the values from Table \ref{tab:materialProperties_UCTest} then yields $\ell = 0.922$ mm. As crack paths are a priori unknown, the problem domain is discretized uniformly into triangular elements, with the characteristic size of element edges set to 0.23 mm. The resulting unstructured mesh contains a total of 112,806 nodes and 224,304 elements. In the matrix region, each node has three active degrees of freedom corresponding to $u_x$, $u_y$ and $\phi$. On the other hand in regions corresponding to inclusions, only the displacement DOFs are active. Consequently the total number of active DOFs is between 293,000 and 298,000 depending on the specific type of boundary conditions used for the phase-field equation.

We conduct two simulation runs to look at what effect the type of phase-field BC imposed at material interfaces has on the predicted behavior of the specimen. As inclusions do not fracture, they are modeled using the numerics class \class{PlaneStrain\_Fe\_Tri3}, which implements the stress equilibrium equation in 2D assuming plane strain conditions. The latter is ``compatible'' with the earlier described numerics class for phase-field fracture with respect to the stress equilibrium implementation, hence they can share the same set of displacement DOFs. However as these two classes are each assumed to have been coded without knowledge of one another, the task of setting up the simulation allowing for their simultaneous use falls to the end user, who must globally declare the number of DOF types to be used and subsequently assign them within the two classes in question so as to be recognized by the latter. This is done purely within the input file as illustrated in Listing \ref{lst:inputFile_UCTest}, with no additional coding required.
\begin{lstlisting}[float,caption={Partial contents of input file for the unconfined compression problem.},label={lst:inputFile_UCTest},stringstyle=\color{black}\ttfamily]
...
*DOF_PER_NODE 3
  1 DofGroup 1 NodalField 1 4 // displacement component, u_x
  2 DofGroup 1 NodalField 2 5 // displacement component, u_y
  3 DofGroup 2 NodalField 3 6 // phase-field, phi
	
*NUMERICS 2
  1 PhsFieldFracture_Fe_Tri3
    NodalDof 1 2 3 // u_x, u_y, phi
    Stage 1
    Subsystem 1 2 // Subsystem number stress equilibrium, followed by subsystem number for phase-field equation
    CellFieldOutput 7
      1 s_xx
      2 s_yy
      3 s_zz
      4 s_xy
      5 ux_x
      6 uy_y
      7 g_xy
    CharacteristicLength 0.922
    CriticalEnergyReleaseRate 2.8e-3
    CrackInitialization No
    
  2 PlaneStrain_Fe_Tri3
    NodalDof 1 2 // u_x, u_y
    Stage 1
    Subsystem 1 // Subsystem number for stress equilibrium
    CellFieldOutput 7
      1 s_xx
      2 s_yy
      3 s_zz
      4 s_xy
      5 ux_x
      6 uy_y
      7 g_xy
		
*MATERIALS 3
  1 Density 0.0
  2 AmorDamageModel
    LinearIsotropicElasticity PlaneStrain 30000 0.25
    QuadraticDegradation
    IrreversibilityThreshold 0.5
  3 LinearIsotropicElasticity PlaneStrain 77000 0.2
	
*DOMAIN_ASSIGNMENTS 2
  "Matrix" Numerics 1 MaterialSet 1 2
  "Inclusions" Numerics 2 MaterialSet 1 3
...
\end{lstlisting}
Likewise, subsystem numbers are assigned for each set of governing equations implemented in the numerics classes. This is necessary so that the right sets of equations are grouped together in the subsequent solution steps. To solve the system of nonlinear equations at each time step, we utilize a solution algorithm implemented in the class \class{AlternateNonlinearMinimization}, which performs a nested cycle of iterations to solve the coupled system of equations: the outer loop alternates between the mechanics and phase-field subsystems, while the inner loop applies the Newton-Raphson algorithm to solve the active subsystem with all inactive unknowns fixed at their values from the previous iteration. On a machine equipped with a single Intel Core i7-8700 CPU (6 cores/12 threads) running 3.20 GHz and with equation assembly, solution and update done in parallel over all available threads, completion of a single iteration in the outer loop that includes one inner iteration for each subsystem takes around 0.9 seconds for above-mentioned discretization.

\begin{figure}
	\centering
	\includegraphics[width=0.4\textwidth]{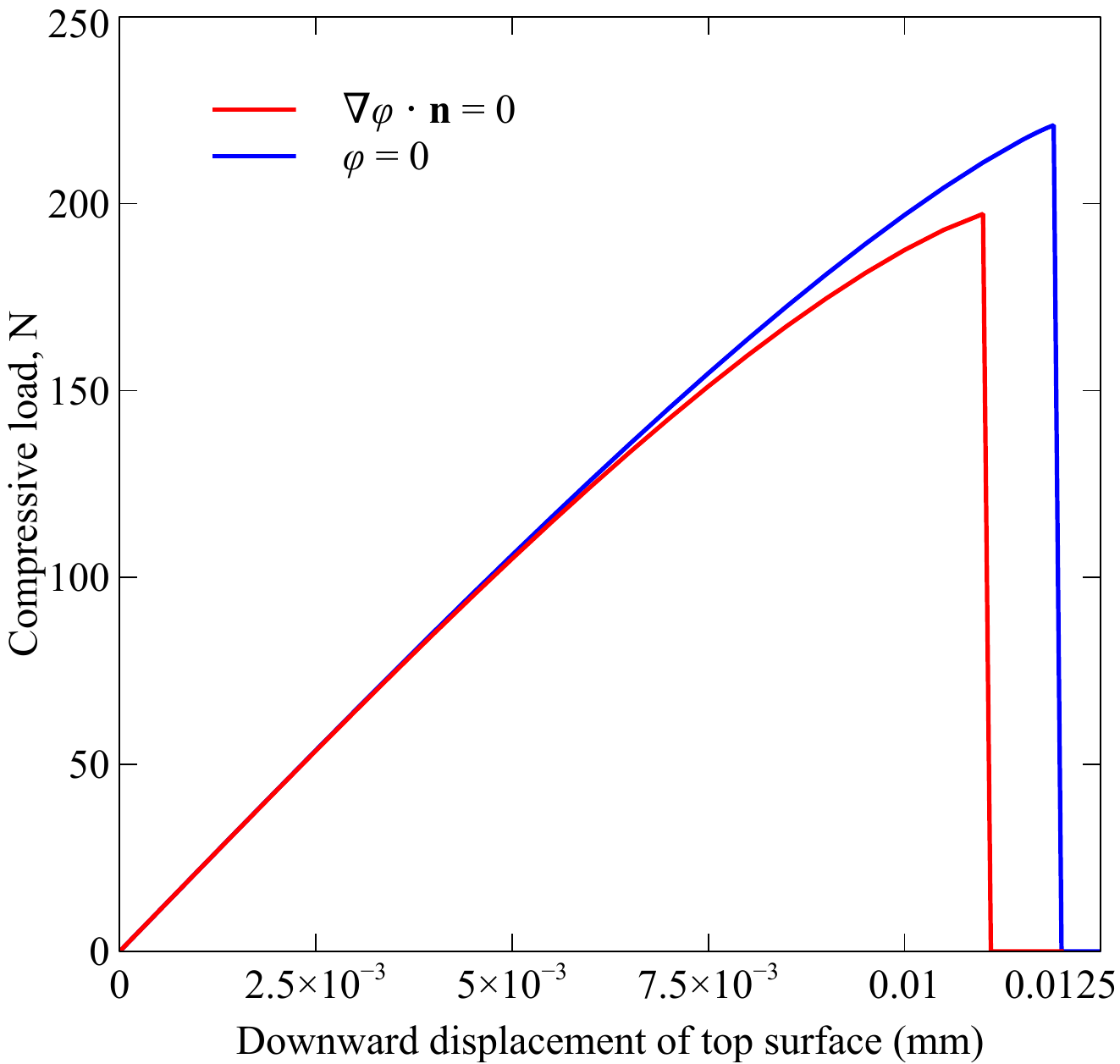}
	\caption{Effect of phase-field boundary condition at material interfaces on load-displacement response for the unconfined compression test.}
	\label{fig:ucTestLoadDisp}
\end{figure}

\begin{figure}
	\centering
	\begin{subfigure}{0.225\textwidth}
		\includegraphics[width=\textwidth]{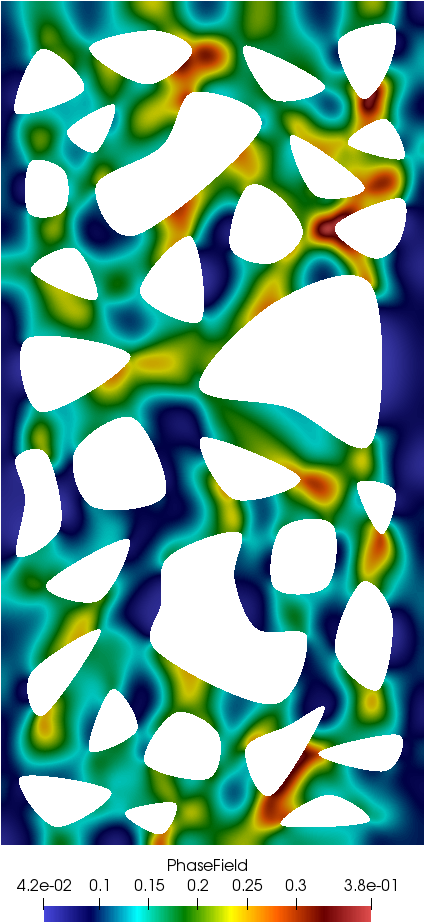}
		\caption{} \label{fig:ucTest_bc1_a}
	\end{subfigure} \hspace{8pt}
	\begin{subfigure}{0.225\textwidth}
		\includegraphics[width=\textwidth]{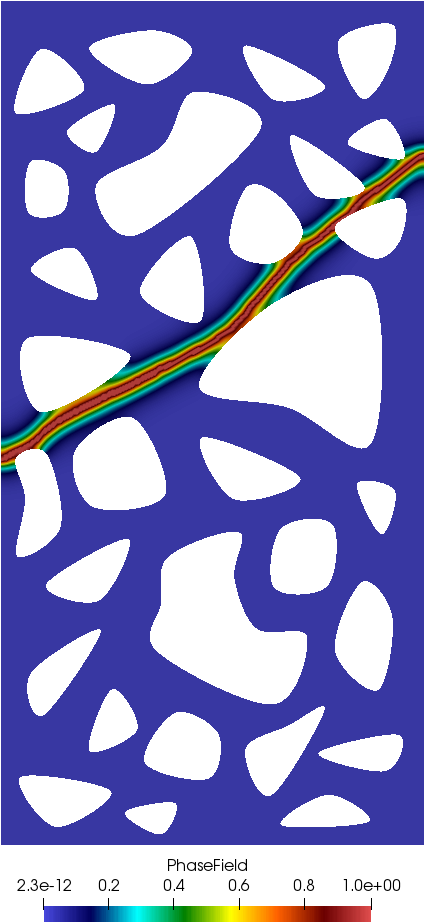}
		\caption{} \label{fig:ucTest_bc1_b}
	\end{subfigure} \hspace{20pt}
	\begin{subfigure}{0.225\textwidth}
		\includegraphics[width=\textwidth]{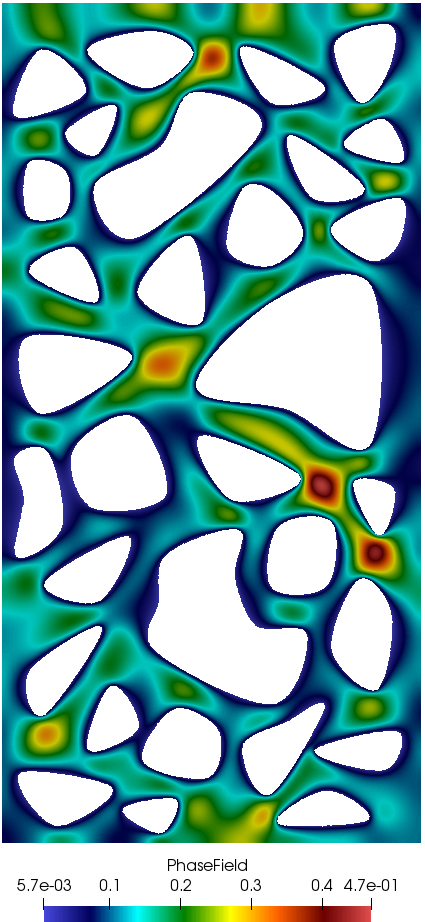}
		\caption{} \label{fig:ucTest_bc2_a}
	\end{subfigure} \hspace{8pt}
	\begin{subfigure}{0.225\textwidth}
		\includegraphics[width=\textwidth]{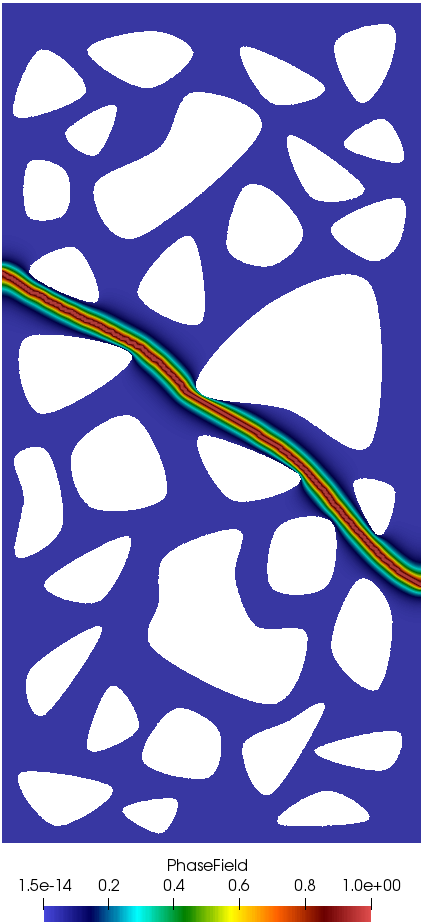}
		\caption{}
	\end{subfigure}
	\caption{Respective phase-field profiles immediately before and after specimen failure for different boundary conditions imposed at material interfaces. (a) -- (b): $\nabla\phi \cdot \mathbf{n} = 0$; (c) -- (d): $\phi = 0$. Note the difference in color scales between (a) and (c), which is necessary in order to sufficiently highlight the areas in which the phase-field attains its peak values prior failure.}
	\label{fig:ucTestCrackPaths}
\end{figure}

\begin{figure}
	\centering
	\begin{subfigure}{0.225\textwidth}
		\includegraphics[width=\textwidth]{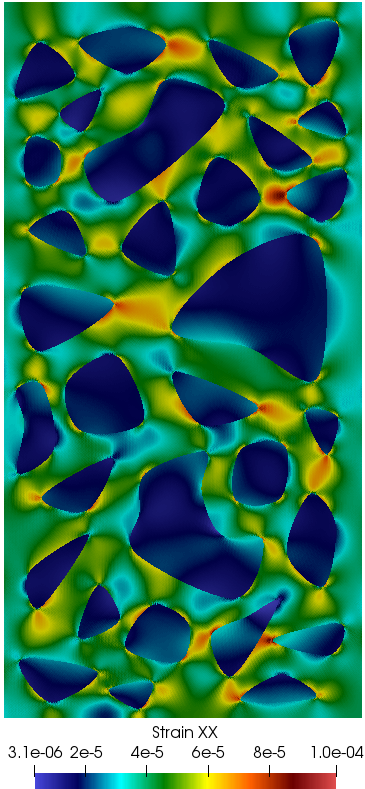}
		\caption{} \label{fig:ucTest_elastic_eps_xx}
	\end{subfigure} \hspace{8pt}
	\begin{subfigure}{0.225\textwidth}
		\includegraphics[width=\textwidth]{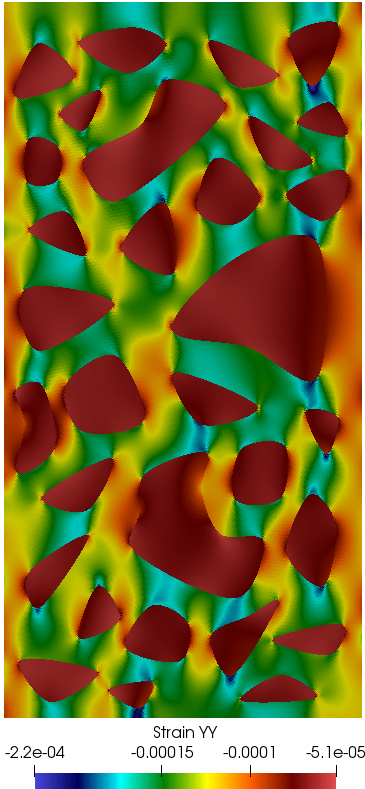}
		\caption{}
	\end{subfigure} \hspace{8pt}
	\begin{subfigure}{0.225\textwidth}
		\includegraphics[width=\textwidth]{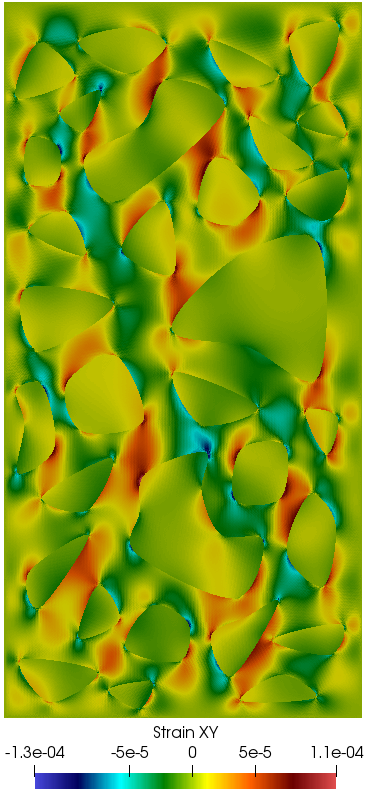}
		\caption{}
	\end{subfigure}
	\caption{Distribution of strain components in the concrete specimen for the unconfined compression test at $U = 0.012$, assuming purely elastic behavior for both aggregate inclusions and cement paste matrix.}
	\label{fig:ucTest_elastic}
\end{figure}

Load-displacement curves for the two simulations are plotted in Figure \ref{fig:ucTestLoadDisp}, where it can be seen that imposing the condition $\phi = 0$ at matrix/inclusion interfaces results in a slightly higher critical load.
Furthermore failure occurs in a brutal manner over a single time step. While both simulations predict the shearing type of failure associated with unconfined compression tests, the resulting crack paths are fundamentally different as shown in Figure \ref{fig:ucTestCrackPaths}.
This difference can be explained by looking at the phase-field profiles just prior to fracture. As can be observed in \ref{fig:ucTest_bc2_a}, imposing $\phi = 0$ at the inclusion/matrix interfaces has the effect of making crack initiation more likely at locations where the inclusions are further apart. This is rather unphysical, since stress concentrations are more likely to occur at interfaces due to the abrupt difference in material parameters. On the other hand, imposing $\nabla\phi \cdot \mathbf{n} = 0$ results in a final crack that ``cuts through'' inclusions (see Figure \ref{fig:ucTest_bc1_b}), arguably violating the crack phase-field theory. Further insight can be gained by running a third simulation wherein both the inclusions and surrounding matrix are modeled as linear elastic. Resulting strain distributions for the latter are displayed in Figure \ref{fig:ucTest_elastic}, where in particular we can observe that the highly damaged regions in Figure \ref{fig:ucTest_bc1_a} correspond to the locations of highest tensile strains ($\epsilon_{xx}$) in the purely elastic simulation (Figure \ref{fig:ucTest_elastic_eps_xx}).
This indicates that the correct crack path for the present problem is more likely the one obtained by imposing $\nabla\phi \cdot \mathbf{n} = 0$ at matrix/inclusion interfaces. We note that such observation is not easily generalized; for instance in the present case it is made within the context that the width of support for the regularized crack that is dependent on the regularization parameter $\ell$ is often larger than the spacing between inclusions which effectively prevents crack initiation when said inclusions are too closely spaced and the condition $\phi = 0$ is imposed at interfaces. A possible remedy is to sufficiently reduce $\ell$ in order to remove such spacing penalization, however this also affects the critical tensile stress at which crack initiation occurs so that further modifications must be made to the fracture model, for instance the use of parametric degradation functions\cite{Sargado2018}. The proper setup and interpretation of fracture phase-field models in the presence of holes and inclusions is currently still an open research topic, and for interested readers the recent work of Nguyen et al.\cite{Nguyen2018} provides a more in-depth discussion. On the other hand our purpose for including the present example in this paper (in addition to those already mentioned) is to demonstrate the usefulness of having all the necessary formulations present in a single code that can then be harnessed by end users according to their needs. Often, a good way to evaluate and gain experience with relatively new and unfamiliar numerical models is to compare their predictions to those obtained through the use of simpler, more understood models. In the current problem for instance, the shift from a fracture model to a purely elastic simulation is done completely via modification of input files, without any additional programming necessary from the end user.

\section{Concluding remarks} \label{sec:conclusion}
In this paper, the design of {\broomstyx}, an efficient object-oriented, multi-method and multiphysics framework has been presented. The code architecture incorporates elements of both framework and library approaches in order deliver equal benefit to both model developers and end users. An important focus of the code is on maximizing flexibility available to end users through input file definitions, which has been demonstrated with several numerical examples. Furthermore, {\broomstyx} implements its own vector and matrix classes along with operator overloading and a lazy evaluation strategy in order to improve code readability without sacrificing efficiency of execution. In order to fully exploit the capabilities of machines having multiple-core processors, the code makes use of shared memory parallelization via OpenMP directives.

At present, the {\broomstyx} framework contains over 25,000 lines of code and serves as the main simulation tool used by the Author for computational research. Robustness of the code for handling different types of problems has been demonstrated in the current work via several numerical examples. In particular we have shown that the code can accommodate models featuring unknowns that are not associated with geometric entities, constraints involving master and slave DOFs as well as the combination of fundamentally different numerical formulations within a monolithic solution scheme. In addition, the framework can handle complicated nonlinear problems that require special solution algorithms. While current implemented models have so far been confined to 2D, the framework readily admits 3D formulations without any modification to the core structure. Extension of the software to allow for distributed memory parallelization through MPI routines is currently a work in progress, along with adaptive remeshing capabilities through the integration of existing libraries. It is hoped that the framework can serve as a platform upon which researchers can implement new models and algorithms in a manner that can be immediately useful to engineers dealing with real world problems, thus eventually speeding up the adoption of such innovations into mainstream industry. To this end, {\broomstyx} is made available as open-source software under Version 3 of the GNU General Public License (GPLv3), and the source code can be downloaded from \url{https://github.com/broomstyx/broomstyx.git}.

\section*{Acknowledgements}
This work was funded by the Research Council of Norway through grant no.\ 228832/E20 and Equinor ASA through the Akademia agreement. I am immensely grateful to Eirik Keilegavlen, Inga Berre and Jan M.\ Nordbotten for taking the time to review the current manuscript and for their many suggestions that have helped to improve its content. I would also like to thank  Robert Kl\"ofkorn for his many insights on advanced C++ programming that have aided me greatly in improving the overall structure of the {\broomstyx} code.
\appendix
\section{Alternate solution of torsion problem involving symmetric systems}
The set of equations given in \eqref{eq:stVenantDiscreteProblem} are asymmetric, however they can be split into two consecutive solutions of symmetric systems. As mentioned in the main discussion, the matrix equation $\mathbf{K} \hat{\mathbf{\omega}} = \mathbf{F}^{\partial\Omega} + \sum_i \mathbf{F}^{\Gamma_i}$ corresponds to a pure Neumann problem so that the coefficient matrix $\mathbf{K}$ is rank-deficient by 1 and $\hat{\mathbf{\omega}}$ can be determined uniquely up to a constant. In such cases it is often the practice to artificially constrain one degree of freedom from the set $\left\{ \hat{\omega}_I \right\}$, however a more elegant way is to introduce a Lagrange multiplier $\lambda$ for the purpose of imposing the constraint
\begin{linenomath}
\begin{equation}
	\int_\Omega \omega \dee\Omega = 0 .
\end{equation}
\end{linenomath}
This results in the symmetric augmented system
\begin{linenomath}
\begin{equation}
	\left[ \begin{array}{cc} \mathbf{K} & \mathbf{Q} \\ \mathbf{Q}^\trp & \mathbf{0} \end{array} \right]
	\left\{ \begin{array}{c} \hat{\mathbf{\omega}} \\ \lambda \end{array} \right\} =
	\left\{ \begin{array}{c} \mathbf{F}^{\partial\Omega} + \sum_i \mathbf{F}^{\Gamma_i} \\ 0 \end{array} \right\},
\end{equation}
\end{linenomath}
which now yields unique values for $\hat{\mathbf{\omega}}$ and $\lambda$. The constants $a_z$, $k_x$ and $k_y$ can subsequently be determined by solving the system
\begin{linenomath}
\begin{equation}
	\left[ \begin{array}{ccc} A & S_x & -S_y \\[0.2em] S_x & I_{xx} & -I_{xy} \\[0.2em] -S_y & -I_{xy} & I_{yy} \end{array} \right]
	\left\{ \begin{array}{c} a_z \\[0.2em] k_x \\[0.2em] k_y \end{array} \right\} =
	\left\{ \begin{array}{c} \mathbf{0} \\[0.2em] -\mathbf{Q}^\trp_x \hat{\mathbf{\omega}} \\[0.2em] -\mathbf{Q}^\trp_y \hat{\mathbf{\omega}} \end{array} \right\} .
\end{equation}
\end{linenomath}
While this method does not necessarily yield the same solution of $\hat{\mathbf{\omega}}$ as \eqref{eq:stVenantDiscreteProblem}, they lead to the same expression for $u_z \left( x,y \right)$, which is after all the relevant quantity describing the cross sectional warping for an arbitrary cross section.

\section{Evaluation of boundary terms on curved edges} \label{app:StVenantBCTerm}
In Section \ref{sec:StVenantTorsion}, a right hand side term must be evaluated for elements having edges that coincide with interior or exterior boundaries of the bar cross section, given by
\begin{linenomath}
\begin{equation}
	F^e_I = \int_{\partial\Omega^e} N^e_I \mathbf{x}_\perp \cdot \mathbf{n} \,\dee\Gamma
	\label{eq:appBCterm}
\end{equation}
\end{linenomath}
As the chosen FE discretization makes use of 6-node isoparametric elements, the element edges over which the above term is calculated (and which in general may be curved) can be treated as 3-node isoparametric 1D elements. For the latter, we assume that a curved element in 2D Cartesian space maps to a master element having nodes at reference coordinates $\xi = -1$, $\xi = 0$ and $\xi = 1$. The spatial coordinates are given by
\begin{linenomath}
\begin{equation}
	\mathbf{x} \left( \xi \right) = \sum_{I=1}^3 N^e_I \left( \xi \right)  \mathbf{x}_I = \left\{ \kern-0.3em \begin{array}{r} x \left( \xi \right) \\ y \left( \xi \right) \end{array} \kern-0.3em  \right\}
\end{equation}
\end{linenomath}
where $N_I$ are the Lagrange shape functions corresponding to a 2nd degree polynomial. On the other hand, $\mathbf{n}$ is given by
\begin{linenomath}
\begin{equation}
	\mathbf{n} \left( \xi \right) = \frac{b}{\sqrt{\mathbf{x}^\prime \left( \xi \right) \cdot \mathbf{x}^\prime \left( \xi \right)}} \left\{ \kern-0.3em \begin{array}{r} -y^\prime \left( \xi \right) \\ x^\prime \left( \xi \right) \end{array} \kern-0.3em  \right\}
\end{equation}
\end{linenomath}
in which $b$ has a value of either $+1$ or $-1$ such that $\mathbf{n}$ points outward from the associated domain element. Finally, $\dee\Gamma$ is given by
\begin{linenomath}
\begin{equation}
	\dee\Gamma = \sqrt{\dee\mathbf{x} \cdot \dee\mathbf{x}} = \sqrt{\mathbf{x}^\prime \left( \xi \right) \cdot \mathbf{x}^\prime \left( \xi \right)} \dee\xi
\end{equation}
\end{linenomath}
Making use of the above expressions and noting that $\mathbf{x}_\perp \left( \xi \right) = \left\{ -y \left( \xi \right), x \left( \xi \right) \right\}^\mathrm{T}$, we can rewrite \eqref{eq:appBCterm} as
\begin{linenomath}
\begin{equation}
	F^e_I = \int_{-1}^1 b N^e_I \left( \xi \right) \left[ y \left( \xi \right) y^\prime \left( \xi \right) + x \left( \xi \right) x^\prime \left( \xi \right) \right] \dee\xi
\end{equation}
\end{linenomath}
As $N_I$ is of degree 2, the expression above involves integration of a 5th degree polynomial. In the code, a 3-point Gauss quadrature is used to obtain exact results.

\bibliographystyle{plain}
\bibliography{Reference.bib}

\end{document}